%% file: main.tex
\documentclass[reprint,floatfix,notitlepage,nofootinbib,twocolumn,superscriptaddress,pra,preprintnumbers]{revtex4-1}
\usepackage[colorlinks=true,citecolor=blue,linkcolor=magenta]{hyperref}
\usepackage{physics}
\usepackage{float}
\usepackage{amsmath}
\usepackage{amssymb}
\usepackage{cancel}
\usepackage{graphicx}
\usepackage{subcaption}
\usepackage{enumitem}
\usepackage{pifont,bbm}
\usepackage{xifthen}
\usepackage{xargs}
\usepackage{multirow}
\usepackage{makecell}
\usepackage{xspace}
\usepackage{pgfplots}
\pgfplotsset{compat=newest}
\usepgfplotslibrary{fillbetween}
\usetikzlibrary{calc,fadings,decorations.pathreplacing,shapes,shapes.multipart,arrows,shapes.misc,intersections,positioning}

\usepackage{simpler-wick}

\usepackage[english]{babel}
\usepackage{xr}
\graphicspath{{./}{./images/}}

\usepackage{ulem}

\DeclareMathOperator*{\argmax}{argmax}

\newcommand{\I}{\mathbb{I}}

\input{tikz_lib}

\makeatletter
\def\bbl@set@language#1{%
  \edef\languagename{%
    \ifnum\escapechar=\expandafter`\string#1\@empty
    \else\string#1\@empty\fi}%
  \@ifundefined{babel@language@alias@\languagename}{}{%
    \edef\languagename{\@nameuse{babel@language@alias@\languagename}}%
  }%
  \select@language{\languagename}%
  \expandafter\ifx\csname date\languagename\endcsname\relax\else
    \if@filesw
      \protected@write\@auxout{}{\string\select@language{\languagename}}%
      \bbl@for\bbl@tempa\BabelContentsFiles{%
        \addtocontents{\bbl@tempa}{\xstring\select@language{\languagename}}}%
      \bbl@usehooks{write}{}%
    \fi
  \fi}
\newcommand{\DeclareLanguageAlias}[2]{%
  \global\@namedef{babel@language@alias@#1}{#2}%
}
\makeatother

\DeclareLanguageAlias{en}{english}

\newcommand{\bw}{BW\xspace}
\newcommand{\stair}{S\xspace}
\usepackage{bm}

\newcommand{\fc}{\hspace{0.1em}\fineq[-0.6ex]{\fill (0,0) circle (0.065);}\hspace{0.1em}}
\newcommand{\ec}{\hspace{0.1em}\fineq[-0.6ex]{\draw (0,0) circle (0.065);}\hspace{0.1em}}
\newcommand{\oS}{\mathcal{S}}

\newcounter{noteinbib}

\begin{document}
\title{Two-stage relaxation of operators through domain wall and magnon dynamics}
\author{Cheryne Jonay$^*$}
\affiliation{Faculty of Mathematics and Physics,
University of Ljubljana, SI-1000 Ljubljana, Slovenia}
\author{Cathy Li$^*$}
\author{Tianci Zhou}
\affiliation{Department of Physics, Virginia Tech, Blacksburg, Virginia 24061, USA}
 
\date{\today}

\begin{abstract}
The out-of-time-ordered correlator (OTOC) is a popular probe for quantum information spreading and thermalization. In systems with local interactions, the OTOC defines a characteristic butterfly lightcone that separates a regime unperturbed by chaos from one where time-evolved operators and the OTOC approach equilibrium. This relaxation has been shown to proceed in two stages. The first stage exhibits an extensive timescale and a decay rate known as the ``phantom eigenvalue", which is slower than the gap of the transfer matrix. In this work, we investigate the two-stage relaxation of the OTOC towards its equilibrium value in various local quantum circuits. We apply a systematic framework based on an emergent statistical model, where the dynamics of two single-particle modes -- a domain wall and a magnon -- govern the decay rates. We show that a configuration with coexisting domain wall and magnon modes generates the phantom rate in the first stage, while competition between these two modes determines the second stage. We also examine this relaxation within the operator cluster picture. The magnon modes translate into a bound state of clusters and the domain wall into a random operator, giving consistent rates. Finally, we extend our findings from random-in-time circuits to a broad class of Floquet models. 
\end{abstract}

\maketitle
\setlength{\skip\footins}{26pt} %
\def\thefootnote{*}\footnotetext{These authors contributed equally.}\def\thefootnote{\arabic{footnote}}

\section{Introduction}
Classical chaos is characterized by its extreme sensitivity to the initial conditions, as illustrated by the ``butterfly effect", where small perturbations can lead to dramatically different phase space trajectories. In contrast, the linearity of quantum evolution precludes exponentially diverging trajectories in the full Hilbert space. However, in recent years, one successful probe of quantum chaos has been found in the form of an out-of-time-ordered correlation function (OTOC)\cite{larkin_quasiclassical_1969,roberts_diagnosing_2015,kitaev2015,roberts_localized_2015, shenker_stringy_2014}, from which a quantum analog of the butterfly effect can be extracted. It has gained traction in studying systems in both high energy and condensed matter physics.  
In certain holographic systems with black holes, the time scales of OTOCs are used to determine the bulk causal structure~\cite{roberts_localized_2015, shenker_stringy_2014,qi_butterfly_2017,shenker_black_2014}. The rate at which the commutator increases is shown to be upper-bounded by the inverse temperature of the system \cite{maldacena_bound_2016,parker_universal_2019,murthy_bounds_2019,liao_nonlinear_2018}, which is saturated in the Sachdev-Ye-Kitaev (SYK) model, a strongly interacting fermionic system in 0+1 dimension~\cite{sachdev_gapless_1993, sachdev_bekenstein-hawking_2015, kitaev2015}. The time scales set by OTOCs and their saturation values are also used to diagnose phases of many-body physics \cite{PhysRevB.107.L020202, PhysRevB.95.165136, PhysRevB.99.224305,huang_out--time-ordered_2016,fan_out--time-order_2016,chen_out--time-order_2016}. More recently, operator spreading has been examined in various random circuit and Hamiltonian models with or without conservation laws~\cite{hosur_chaos_2016,xu_accessing_2018,zhou_levy_flight_2020,xu_locality_2019, Khemani_operator_hydro_2018,2018PRX_Rakovszky,vonKeyserlingk2018operator,chen_operator_2018,gu_energy_2017,nahum2018operator,lucas_quantum_2019,aleiner_microscopic_2016,bertini_scrambling_2020-1}, revealing how the relaxation of conserved quantities introduces additional timescales for the spread of quantum chaos. %

The fundamental mechanism underlying the growth (or decay) of the OTOC is scrambling - the process by which quantum information spreads through interactions. This is often studied in the operator language, by tracking how an initially localized operator grows in time. %
To understand this precisely, we consider a spin chain and define the OTOC of two initially local operators $V$ and $W$ as the following infinite temperature average
\begin{align}
\label{eq:otoc_intro}
\text{OTOC}(x,t) = \langle V(0, t) W(x, 0) V(0,t) W(x,0) \rangle 
\end{align}
where $\langle \cdots \rangle = \frac{1}{q^L}\rm Tr(...)$. The operators $W$ at position $x$ and $V$ at position $0$ appear in a non-time ordered sequence, hence 'out-of-time-ordered' correlator. This %
allows us to compare two quantum states: one first perturbed by $W(x,0)$ then by $V(0,t)$, and another first perturbed by $V(0,t)$ then by $W(x,0)$. At $t=0$, the local operators $W(x,0)$ and $V(0,0)$ are spatially separated and commute, resulting in identical states. Consequently, $\text{OTOC}(x,0)$ is a time-independent $\mathcal{O}(1)$ constant. However, as the operator $V(0,t)$ evolves and spreads, it ceases to commute with $W(x,0)$, causing the overlap between these differently evolved states to decay. This decrease in overlap quantifies how much the $W(x,0)$ perturbation affects the system when applied before or after $V(0,t)$. The operator $V(0,t)$ typically grows outward at a characteristic ``butterfly velocity" $v_B$, making the OTOC a probe of quantum analog of the butterfly effect\cite{shenker_black_2014,kitaev2015}. The time at which $V(0,t)$ has scrambled sufficiently to cause an $\mathcal{O}(1)$ decrease in the OTOC is known as the scrambling time\cite{shenker_black_2014,hosur_chaos_2016,susskind_switchbacks_2014}. 

\begin{figure}[h]
\centering
\includegraphics[width=\columnwidth]{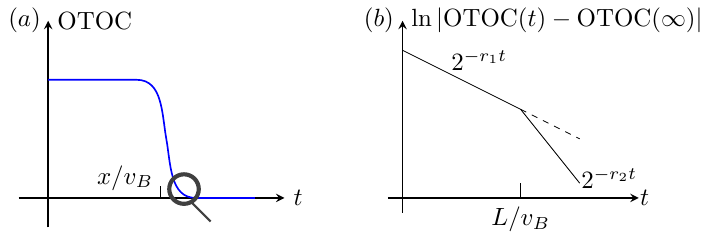}
\caption{(a) The OTOC is initially small and experiences a drastic growth at a time $\sim x / v_B$, and then saturates to an equilibrium value. We focus on how it converges to the saturation value after the time scale of $\sim x / v_B$. (b) The OTOC can show two-stage decay with rates $r_1$ and $r_2$ ($r_1$ possibly < $r_2$), where $L$ is the system size. %
}
\label{fig:otoc_schematics}
\end{figure}

Our work focuses on how the OTOC converges to the steady-state value at long times beyond the initial operator spreading and how the OTOC relaxes to its saturation value. We consider models of local quantum circuits with a well-defined $v_B$. When $t \ge \mathcal{O}(x / v_B)$, the operator $V(0,t)$ becomes close to completely random at the location of $W(x,0)$, which gives the saturation value of the OTOC. 
Interestingly, in a finite system, the relaxation process takes place in two stages, each characterized by different decay rates~\cite{znidaric_phantom_2023}. In both random and non-random circuits, the relaxation in the first stage does not necessarily follow the gap of the %
transfer matrix of the OTOC evolution. Instead, the OTOC %
decays at the rate of ``phantom" eigenvalue\cite{znidaric_phantom_2023} that lies within the actual gap of the transfer matrix spectrum. The duration of this first stage is characterized by a time scale of $L / v_B$, which indicates a failure to exchange the thermodynamic limits of $L\rightarrow \infty$ and $t\rightarrow \infty$.

The emergence of the phantom eigenvalue and the two-stage decay have also been observed in the relaxation of other quantities, the half-state purity being one of the earliest examples~\cite{bensa_fastest_2021, bensa_purity_2022, bensa_two-step_2022, jonay_physical_2024,znidaric_phantom_2023,znidaric_solvable_2022}. To briefly recap, the half-system purity decays exponentially during a quantum quench with local interactions. This is consistent with the linear growth of entanglement entropy before the saturation time $\mathcal{O}(L/v_E)$, where $v_E$ is the entanglement growth rate. However, after $\mathcal{O}(L/v_E)$, the purity can decay at a different rate, which can be slower or faster. In the case when the second stage decay is faster, the first stage decay is precsiely the phantom eigenvalue. %
Our previous work~\cite{jonay_physical_2024} also identified a similar transition for the one-point correlation functions (to be discussed in Sec.~\ref{sec:cluster} in the cluster picture), as well as the relaxation of entanglement spectra toward its equilibrium distribution.

There are generally two approaches for understanding the emergence of the phantom eigenvalue. The first involves the analysis of the pseudo-spectrum of a non-hermitian transfer matrix~\cite{znidaric_solvable_2022, znidaric_phantom_2023}. The left and right eigenstates of a non-hermitian transfer matrix are mutually bi-orthogonal, but they are not necessarily simultaneously normalized. The norm of the left and right eigenvectors can be exponentially large in $L$ but constant in $t$ . The norm remains large when $t \ll L$, which shadows the true decay before $t \sim L/ v_B$ and gives rise to a phantom decay. This is an empirical picture explaing the phantom eigenvalue of the purity through the norm of the left and right eigenvector. 

We adopted a different perspective~\cite{jonay_physical_2024} and interpret the relaxation rates as the free energies of two emergent modes in the effective purity dynamics - a magnon and a domain wall~\cite{nahum_quantum_2017,zhou_entanglement_2020,zhou_nahum_emergent_stat_mech2018,vonKeyserlingk2018operator,jonay2018coarsegrained}, (Sec.~\ref{subsec: mode}).This physical fremework precisely predicts the asymptotic rates of both stages and reveals a mechanism of geometric and dynamical nature leading to the rich phenomenology of two-stage thermalization.

In this work, we extend our physical theory of two-stage decay to describe the relaxation of the OTOC. We find that the two modes -- the domain wall and magnon -- are sufficient to capture the trajectories of the system towards its saturation value. Our main result shows that the boundary conditions imposed by the OTOC in the effective model create a scenario where the first stage hosts a coexisting domain wall and magnon mode whose decay rate is a fraction of the magnon rate. %

The spreading of operators was initially described using the cluster picture, where Pauli strings are represented as expanding clusters under time evolution\cite{nahum2018operator,vonKeyserlingk2018operator,nahum_real-time_2022-1,chen_operator_2018,qi_quantum_2018,roberts_operator_2018}. We apply the same framework to understand the relaxation of the OTOC. In this language, what we refer to as a magnon in the spin picture becomes a bound state of clusters whose decay rate can be directly read off from the two-body transfer matrix. Thus, this basis is particularly effective for predicting scenarios where magnon rates dominate. The domain wall corresponds to a random operator that is a superposition of occupied and unoccupied clusters. This viewpoint helps interpret the reverse transition observed in one-point functions. However, since the domain wall represents a superposition of states in the cluster picture, the classical configurations in OTOC relaxation are less intuitive. 
We conclude that the cluster picture is not a suitable basis when the domain wall dominates.

\tableofcontents

\section{Quantum Circuits Setup and Existing Results}
\label{sec:review}

We first review the setup of local quantum circuits and known numerical results of the two-stage decay of the OTOC. For clarity, we begin with the mathematical formulation. The OTOC can also be expressed  as a term inside the out-of-time-ordered commutator:
\begin{align}
\label{eq: otoc}
&\text{OTO-Commutator}(x,t) = -\frac{1}{2} \langle [ W(x,0), V(0,t]^2 \rangle \\
=&-\langle W(x,0)^2 V(0,t)^2 \rangle + \langle (W(x,0) V(0,t))^2 \rangle . \notag
\end{align}
In this work, $\langle \cdots \rangle$ in OTOC is the expectation value in an infinite temperature ensemble. The first term is a time-ordered correlator of $W(x)^2$, and $V^2(0)$ evolved to time $t$, and generally dissipates to its steady-state value in an $\mathcal{O}(1)$ time scale. As a result, the OTO-Correlator and OTO-Commutator share the same relaxation behaviors for our interested time regime $t> x /v_B$. In this work, we reserve the acronym OTOC for the OTO-Correlator defined in Eq.~\ref{eq:otoc_intro}. 
We will set the displacement of the two operators $x \ll L$ to sidestep the scrambling physics before $x /v_B$. We use 
\begin{equation}
\label{eq:relaxation}
    |\text{OTOC}(x,t) - \text{OTOC}(x,\infty)|
\end{equation}
to quantify the relaxation process. 

\begin{figure}[h]
\centering
\includegraphics[width=\columnwidth]{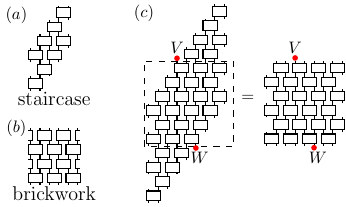}
\caption{The local quantum circuits we consider have two-qudit gates stacked as a (a) staircase or (b) brickwork geometry. (c) For an open boundary condition, the OTOCs are equivalent as we can freely remove or add unitaries outside the light cones of both operators without affecting the OTOC.}
\label{fig:bw_s}
\end{figure}

We model the local interactions via a quantum circuit consisting of nearest-neighbor gates. The gates are stacked to form a brickwork (BW) or staircase (S) geometry as shown in Fig.~\ref{fig:bw_s}(a) and (b)\cite{bensa_fastest_2021}. 

In the brickwork structure, adjacent layers of gates are offset by one lattice spacing, alternating between even and odd layers. We define one time step as a single layer (either even or odd), differing from the transfer matrix formalism where two consecutive layers constitute one time step to preserve time translation invariance. For the staircase geometry, we notice that the interior of the tensor network is no different from the brickwork geometry, and we use the same unit of time convention. This configuration can be experimentally realized through sequential gate applications, known as a "sequential circuit" in quantum computing terminology\cite{Schoen_et_al_seq,anikeeva_recycling_2021}. 

A general two-qubit gate can always be parameterized as follows\cite{khaneja_cartan_2000,kraus2001optimal}
\begin{equation}
\label{eq:cartan-u}
u = \fineq[-0.8ex]{
\sufour[u_{\rm sym}][u_1][u_2][u_3][u_4]}\,.
\end{equation}
where $\{u_i\}_{i=1}^4$ are single-qubit unitaries. In most of this work, we sample the $\{u_i\}_{i=1}^4$ according to the Haar measure~\cite{bensa_two-step_2022} and write  $u_{\rm sym}$ in terms of the commutative Pauli generators 
\begin{equation}
\label{eq:u_sym}
    u_{\rm sym} = \exp{\left[ i \frac{\pi}{4} (a_x X\otimes X + a_y Y\otimes Y + a_z Z\otimes Z) \right]},
\end{equation}
where $X$, $Y$, $Z$ are Pauli matrices. 
The resulting ensemble is parametrized by $(a_x, a_y, a_z)$. When any two of the three numbers are $\pm 1$, the gates become dual unitary, i.e., they are unitary both in the spatial and temporal directions\cite{Bertini_2019correlations,Bertini_2018sff,Claeys_2020,Gopalakrishnan_2019,gutkin2020local,Aravinda2021,Prosen2021roundaface,Zhou_Harrow_2022,ho_exact_2022}. The quantum dynamics becomes tractable in the limit of dual unitarity, where the decay rate of observables can be described by the decay rate of a domain wall $r_{\rm DW} = 1$ \cite{zhou_nahum_entanglement_membrane2019,Foligno_Bertini_2023,Zhou_Harrow_2022,Claeys_2020,Bertini_2019entanglement} and that of a magnon the $r_{\rm mag} = \ln \frac{3}{2-\cos( \pi a_z )}/\ln (2)$ \cite{jonay_physical_2024}.

Tab.~\ref{tab:bz_otoc_summary} includes the two-stage decay rates~\cite{bensa_two-step_2022} found in the case of $(a_x, a_y, a_z) = (1,1,0.5)$ for open boundary condition (OBC) and periodic boundary condition (PBC).
\begin{table}[h]
\centering
\begin{tabular}{ |c|c|c| } 
 \hline
     & $r_1$  & $r_2$\\\hline
    \bw , OBC & $\frac{1}{2}\ln \frac{3}{2} /\ln 2$  & $\ln \frac{3}{2}/\ln 2$  \\\hline    
    \bw, PBC & $\frac{1}{2} \ln \frac{3}{2}/\ln 2$ & $\ln \frac{3}{2}/\ln 2$  \\\hline
    \stair , OBC  & $\frac{1}{2} \ln \frac{3}{2}/\ln 2$ & $\ln \frac{3}{2}/\ln 2$  \\\hline
    \stair , PBC  & $\frac{1}{2} \ln \frac{3}{2}/\ln 2$ & $\frac{1}{2} \ln \frac{3}{2}/\ln 2$ \\\hline
\end{tabular}
\caption{The decay rates of the OTOC for the $(1, 1, 0.5)$ circuit displayed in Ref.~\cite{bensa_two-step_2022} (translated to our units). The magnon rate for this circuit is $\ln \frac{3}{2}/(\ln(2) )$.}
\label{tab:bz_otoc_summary}
\end{table}

The number $\ln \frac{3}{2}/\ln 2$ rounded from numerics is conjectured to be $\ln \frac{3}{2-\cos( \pi a_z ) }/ \ln 2$ for general $a_z$,  which is the second largest eigenvalue of the transfer matrix. For all cases, the first stage decay is half of this value, implying the existence of a ``phantom eigenvalue".  

The half system purity of the dual unitary circuit $(1,1, a_z)$ also exhibits two-stage decay with the first stage decay rate $r_1$ always equal to $1$, and the second stage rates given by the following function of $a_z$: %
\begin{equation}
\label{eq:two_rates}
    r_2 = \begin{cases}
        1 \quad &  a_z \le \frac{1}{3} \\
        \ln \frac{3}{2-\cos( \pi a_z ) }/ (\ln (2) ) 
        \quad & a_z \ge \frac{1}{3} \\
    \end{cases}.
\end{equation}
In our previous work, we have shown that $r_2=1$ is the decay rate of a domain wall and $r_2=\ln \frac{3}{2-\cos( \pi a_z ) }/ (\ln (2) )$ is the rate of a magnon. The construction of these modes will be reviewed later in Sec.~\ref{subsec: mode}. Their competition result in the different behaviors on the two sides of $a_z = \frac{1}{3}$.

We conclude this section by showing that the BW and S geometry are equivalent for the OTOC when the boundary condition is open. Fig.~\ref{fig:bw_s}(c) shows that by moving unitaries outside the causal light cones of the two operators, the possible trajectories between the operators, and hence the OTOC, can remain invariant. Therefore, the decay rates in the two circuits must be identical. However, the equivalence breaks down when the boundary condition is periodic since it identifies the points on the light cone in the S-geometry.

\section{The Effective Modes of the Emergent Magnets}

In this section, we review and develop the emergent statistical model of random quantum circuits. The mapping between the quantum circuits and the classical model is achieved through the quench average of the unitary gates, denoted by an overline $\overline{f(U)} = \int dU f(U)$. We realize the global time evolution $U(t)$ through the \bw or \stair geometry as in Fig.~\ref{fig:bw_s} (a)-(b). Since the time evolution is random in space and time, the ensemble average factorizes, and can be performed locally on the  
$\{u_i\}_{i=1}^4$ in Eq.~\ref{eq:cartan-u} at each spacetime point.

\subsection{The spin model and boundary conditions imposed by OTOC}
The dynamics of OTOC in random quantum circuits can be understood through two equivalent pictures: One is associated with operator spreading, in which the growing support of the %
operator %
is captured through the cluster basis\cite{nahum_quantum_2017,zhou_entanglement_2020,zhou_nahum_emergent_stat_mech2018,vonKeyserlingk2018operator,jonay2018coarsegrained} (Sec.~\ref{sec:cluster}). The other is through an emergent magnet with classical Ising degrees of freedom. The latter is familiar from the context of studying entanglement dynamics\cite{vonKeyserlingk2018operator,2018PRX_Rakovszky,zhou_entanglement_2020,nahum_quantum_2017,vasseur_entanglement_2019,harrow_church_2013,liao_effective_2022,hunterjones_unitary_designs_from_stat_mech_2019,brown_convergence_2010,arnaud_efficiency_2008}. 

Here we review the setup in terms of the Ising degrees of freedom~\cite{chan_solution_2018,nahum_quantum_2017,zhou_nahum_emergent_stat_mech2018,zhou_entanglement_2020,vonKeyserlingk2018operator,Vasseur_holographicTN_2019,Jian_mi_criticality_2020,Bao_theory_mipt_2020,hunterjones_unitary_designs_from_stat_mech_2019,Liu_ee_gravity_2021,Fisher_Vijay_CircuitsReview,jonay_physical_2024,Li_dissipation_PRB}. The microscopic rules of the statistical model (or the stochastic process) are discussed in Ref.~\cite{jonay_physical_2024}. Below we set up the graphical notations and clarify the differences between the OTOC and the purity. 

\begin{figure}
    \label{fig:schematic_otoc}
    \centering
\includegraphics[width=0.95\linewidth]{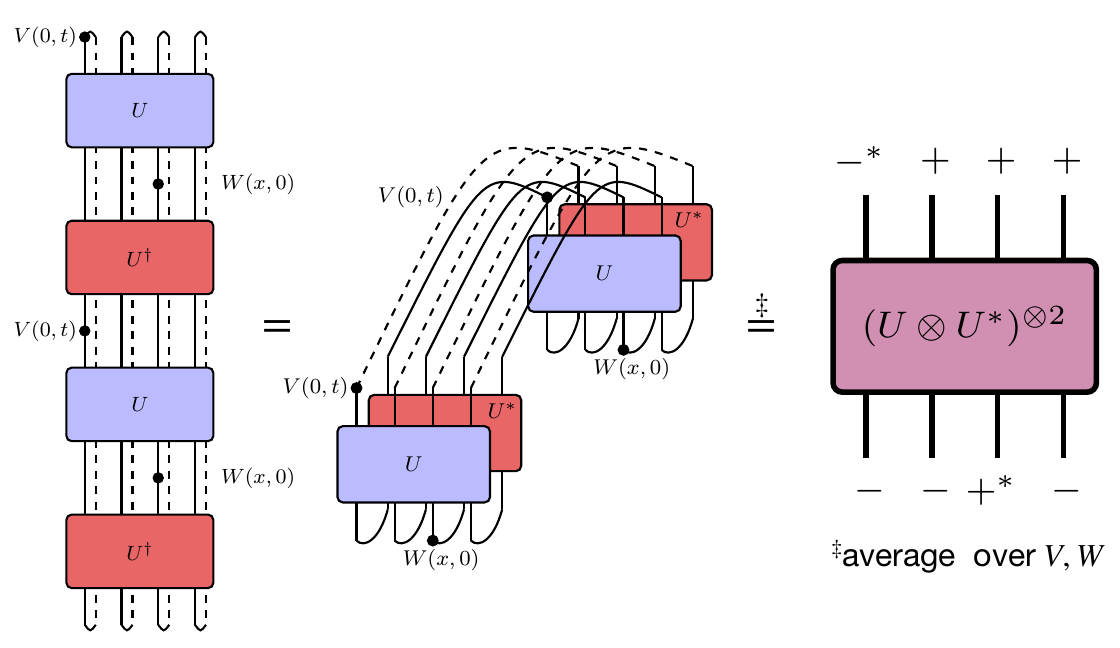}
    \caption{Sketch of the folding procedure of the out-of-time-ordered correlator and mapping of the boundary contractions to effective degrees of freedom. $\ket{+}$ is akin to two stacked rainbows \eqref{eq: pairing +}, $\ket{-}$ to two adjacent ones \eqref{eq: pairing -}. Random traceless operator insertions at time $t$ result in $\ket{-^*}$ \eqref{eq: otoc bc V}, while random traceless operator insertion at $t=0$ results in $\ket{+^*}$ \eqref{eq: otoc bc W}. In this picture, we have not yet illustrated how the \textit{bulk} degrees of freedom get mapped to effective ones. This requires taking a Haar average as in \eqref{eq: u in +-}. 
    }\label{fig:schematic_otoc}
\end{figure}

The unitary matrix and its conjugation appear twice as $u \otimes u^* \otimes u \otimes u^*$ in the diagram of the OTOC (Fig.~\ref{fig:schematic_otoc}). After Haar random averaging, the dominant contribution comes from particular pairings of the unitary and its conjugate, while other contributions cancel~\cite{collins_integration_2006}. The two ways to pair $u$ and $u^*$ are as follows:
\begin{align}
\ket{+}&:= \,\, \wick{
    \c u \otimes  \c u^* \otimes  \c u \otimes  \c u^*  
  } \label{eq: pairing +}\\
\ket{-}&:= \,\, \wick{
    \c2 u \otimes  \c1 u^* \otimes  \c1 u \otimes  \c2 u^* 
  }.\label{eq: pairing -}
\end{align}
These pairings define the two Ising spins $\{\ket{\pm} \}$. They are the only degrees of freedom that survive the random averaging considered in this paper.  Since the Ising basis is not orthonormal, with overlaps $\braket{+}{-} = \braket{-}{+} = q$ and $\braket{+}{+} =\braket{-}{-}  = q^2$, we introduce the dual basis,
\begin{align}
\label{eq: dual_state}
    \ket{+^* } = \frac{1}{q^2 -1} \left( \ket{+} - \frac 1 q \ket{-} \right) \\
\label{eq: dual_state_2}
    \ket{-^*} = \frac{1}{q^2 -1} \left( \ket{-} - \frac 1 q \ket{+} \right).
\end{align}
This enforces orthonormality as $\braket{+^*}{-} = \braket{-^*}{+} = 0$ and $\braket{+^*}{+} = \braket{-^*}{-} =1$, and is convenient when extracting the dominant mode from superposition states. The averaged unitaries and their conjugates can then be written as a projector to the 2-dimensional space of the Ising spins:
\begin{align}
    \overline{u\otimes u^* \otimes u \otimes u^*} = \ket{+} \bra{+^*} + \ket{-} \bra{-^*}.
    \label{eq: u in +-}
\end{align}
The dynamics of the new averaged theory admit two interpretations: it can be viewed either as a Markovian process on the state space $\mathbb{R}_2^{\otimes L}$ spanned by ${\ket{\pm} }$ at each site, or as a two-dimensional partition function of an Ising model whose transfer matrix is the transition matrix of the Markovian process. We will use two terminology interchangeably. 

For a two-qudit Haar random gate acting on a Hilbert space of dimension $q$, we obtain a $4 \times 4$ transition matrix $M$ after random averaging. In the spin basis $(|--\rangle, |-+\rangle, |+-\rangle, |++\rangle )$, the elements are 
\begin{equation}
    M_{\rm Haar} = 
    \begin{pmatrix}
    1 & h & h & 0\\
    0 & 0 & 0 & 0\\
    0 & 0 & 0 & 0\\
    0 & h & h & 1
    \end{pmatrix}. 
\label{eq: Haar transition mat}
\end{equation}
Notice that the possible transitions are restrictive: unitarity dictates $|+ + \rangle$ and $|-- \rangle$ to be invariant, while $|+-\rangle$ and $|-+ \rangle$ can transition to $(|++\rangle + |--\rangle)$ with equal weight $h = \frac{q}{q^2 + 1}$. This \textit{domain wall} mode is the only permissible mode in Haar random dynamics.

In contrast, the more general random two-qubit gates parameterized by $(a_x, a_y, a_z)$ give rise to the following transition matrix after single-qubit averaging: 
\begin{equation}
\label{eq: transition mat}
    M(a_x, a_y, a_z)  = 
    \begin{pmatrix}
    1 & h & h & 0\\
    0 & b_+ & b_- & 0\\
    0 & b_- & b_+ & 0\\
    0 & h & h & 1
    \end{pmatrix},
\end{equation}
where $h=(3-v)/9$, $b_{\pm} = (3 \pm 6u + 5v)/36$ and $u = \cos ( \pi a_x) + \cos ( \pi a_y ) + \cos ( \pi a_z) $, $v = \cos( \pi a_x) \cos( \pi a_y ) + \cos( \pi a_y) \cos( \pi a_z ) + \cos( \pi a_z) \cos( \pi a_x )$~\cite{bensa_fastest_2021}. This process admits more transitions, such as
\begin{align}
    \ket{+-} \rightarrow h \ket{++} + h\ket{--} + b_{+} \ket{+-} + b_{-} \ket{-+}. 
\end{align}
Particularly, the new swap rule $\ket{+-} \rightarrow \ket{-+}$, weighted by amplitude $b_{-}$, allows for a new mode which we call {\it magnon}. It is distinct from the {\it domain wall} mode; we formally define and distinguish them in Sec.~\ref{subsec: mode}. 

The boundary sites at times $t$ (top) and $0$ (bottom) without operator insertion are occupied by $\langle + |$ and $|- \rangle$ respectively, as shown in Fig.~\ref{fig:schematic_otoc}. For the sites with operator insertions, the single-site unitary invariance of the ensemble allows for an additional random average of the operators. When $V$ and $W$ are traceless, the boundary conditions simplify to:
\begin{align}
&\overline{\fineq[-.4ex][0.6][0.6]{\oidst[0][0][$V$][$V$][r]}} = \Tr( V^2 )|-^* \rangle \label{eq: otoc bc V} \\
&\overline{\fineq[-.4ex][0.6][0.6]{\oswapst[0][0][$W$][$W$][r]}}= \Tr( W^2 )|+^* \rangle \label{eq: otoc bc W}.
\end{align}

We now combine the transfer matrix and the boundary condition to compute the OTOC. The transition matrix at time $t$ applied to the entire system $ {\mathcal{M}}(t)$ is given by a tensor product of the $4 \times 4$ transition matrix $M$ associated with the two-site gate. The OTOC can now be expressed solely in terms of the transition amplitudes of a stochastic process. The initial and final states are the boundary conditions for the statistical mechanical model, specified by \eqref{eq: otoc bc V} and \eqref{eq: otoc bc W}: 
\begin{align}
   &\overline{\langle W(x,0) V(0,t) W(x,0) V(0,t) \rangle} =\Tr(V^2) \Tr(W^2) 
   \frac{1}{q^{L}}\nonumber \\
   &\times 
     \bra{- \cdotp \cdotp - +^* -\cdotp \cdotp -} \prod_{t'=1}^t\mathcal{M}(t') \ket{-^* + \cdotp \cdotp +} .
\end{align}
Here, the dual basis $\ket{-^*}$ and $\ket{+^*}$ denotes the location of the operator insertions $V$ and $W$. 

The boundary condition can be further simplified. The initial state in \eqref{eq: otoc bc V} can be expanded via Eq.~\eqref{eq: dual_state_2}
\begin{equation}
\ket{-^* + \cdotp \cdotp +} 
 = \frac{1}{q^2-1} \ket{- + \cdotp \cdotp +}  -\frac{1}{(q^2-1)q}  \ket{+ + \cdotp \cdotp +}. \label{eq: magnon_timeT}
\end{equation}
The second term can be dropped: The all -$\ket{+}$ state is invariant under the action of ${\mathcal{M}}$, and only contributes to a time-independent constant to the OTOC. This is irrelevant for quantifying the decay. The last simplification we make in defining the partition of the OTOC is the removal of the intensive constants multiplying the amplitudes,%
\begin{equation}
\begin{aligned}
\label{eq: otoc_Z}
&Z_{\rm OTOC}(t) = \cancelto{\text{set to }1}{\Tr(V^2) \Tr(W^2) 
   \frac{1}{q(q^2 -1 )}}\\
&\frac{1}{q^{L-1}}
    \bra{- \cdotp\cdotp - +^* -\cdotp \cdotp -} \prod_{t'=1}^t \mathcal{M}(t') \ket{- + \cdots +}. 
\end{aligned}
\end{equation}
$Z_{\rm OTOC}(t)$ can be viewed either as a partition function of a spin model or a weighted amplitude of the Markov process defined by the overlap with the state $\bra{-\cdotp - +^* - \cdotp -}$.
In our notation, a ket $\ket{\cdot}$ represents the fixed top boundary of the spin model and the initial condition of the Markov process, and a bra $\bra{\cdot}$ represents the bottom boundary and the state whose overlap with output of the Markov process we are interested in. In the terminology of Sec.~\ref{subsec: mode}, the bottom boundary will pick out a $\bra{+}$ magnon from the bulk by the orthogonality condition of the dual basis.

\subsection{Emergent single-particle modes}
\label{subsec: mode}

In a system with time translation symmetry, the relaxation of a physical observable is generally determined by the gap of the corresponding transfer matrix. When the transfer matrix is non-hermitian, a ``phantom eigenvalue" within the gap can appear\cite{znidaric_phantom_2023}, which determines the relaxation of the first stage (up to times proportional to $L$)~\cite{bensa_fastest_2021}.

In our previous work~\cite{jonay_physical_2024}, we found two important single-particle modes in the effective statistical model of the purity dynamics, whose competing free energy determines the two-stage decay rates.%
One is the domain wall mode, which is the region between a contiguous $+$ and $-$ domain.
\begin{figure}
\centering 
\includegraphics[width=0.6\columnwidth]{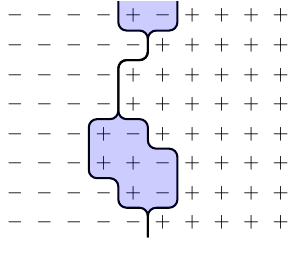}
\caption{An illustration of a domain wall, which separates domains of $-$ and $+$. A domain wall can have a finite width (blue region). In this example, it is the microscopic region separating the first $+$ and the last $-$ counting from the left.}
\label{fig: DW scheme}
\end{figure}

In the Markovian picture of the purity dynamics, the top boundary condition is a domain wall that always dominates in the first stage. In a chaotic quantum circuit, the width of the domain wall is suppressed by chaos, even in systems without random averaging. The bubbles in Fig.~\ref{fig: DW scheme} are not necessarily dense but have $\mathcal{O}(1)$ length (and also $\mathcal{O}(1)$ width by causality). The single particle nature of the thin domain wall allows us to describe its free energy by a single line tension function~\cite{zhou_entanglement_2020,jonay2018coarsegrained}.

\begin{figure}
    \centering
    \includegraphics[width=0.6\columnwidth]{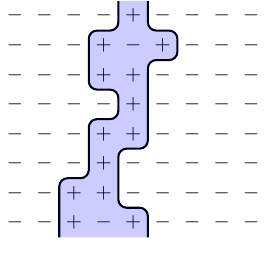}
    \caption{An illustration of a magnon, defined as a region between two domains of the same spin. The illustration shows a $+$ magnon, which is the microscopic region between the first $+$ and last $+$ starting from the left. Similarily, a $-$ magnon seperates the neighboring domains of $+$. %
    }
    \label{fig: magnon scheme}
\end{figure}

The other mode is a magnon, defined by the region of spins sandwiched between domains of the same spin, e.g., Fig.~\ref{fig: magnon scheme}. For similar reasons as in the domain wall example, %
the width of a magnon is suppressed to be $\mathcal{O}(1)$, so a magnon is a bound state of two domain walls separated by an $\mathcal{O}(1)$ distance. 
In a dual unitary circuit ($a_x = a_y = 1$), the domain wall free energy per unit time is $1$\cite{Foligno_Bertini_2023,Bertini_2019entanglement,zhou_nahum_entanglement_membrane2019,Zhou_Harrow_2022}, while the magnon rate is\cite{jonay_physical_2024}
\begin{equation}
    r_{\rm mag} = \frac{\ln \frac{3}{2-\cos( \pi a_z )}}{\ln (2)},
\end{equation}
which can be computed through the quantum channel on the light cone.

\subsection{The fantastic trajectories and where to find them}
\label{sec: trajectories}
To understand the decay rates, we need to understand the dominant modes of the statistical mechanics model. This is the topic of this section. %

We begin with the top boundary at time $t$. The initial magnon state $\ket{- + \cdots +}$ will evolve to a superposition of configurations $|z(t)\rangle$, which can be expanded as 
\begin{align}
  |z(t) \rangle &\equiv \prod_{t'=1}^{t} \mathcal{M}(t') | - + \cdots + \rangle
  = \hspace{-.2cm} \sum_{z \in \{+,-\}^{\otimes L}} \hspace{-.2cm} C_z(t) | z \rangle. 
\end{align}
For our transfer matrix $\mathcal{M}$, the coefficients $C_z(t)$ are real but not necessarily positive. The OTOC is the overlap between the evolved state $\ket{z(t)}$ and the bra state at the bottom boundary, which can be written as a weighted sum of the coefficients,
\begin{align}
Z_{\rm OTOC}(t) &= \bra{- \cdots - +^* - \cdots -}z(t) \rangle \nonumber\\
&= \hspace{-.3cm}  \sum_{z \in \{+,-\}^{\otimes L}} \hspace{-.2cm} C_{z} (t) \langle - \cdots - +^* - \cdots -|z \rangle\nonumber \\
&= \hspace{-.3cm}  \sum_{\substack{z_0 = +, \\ z \in \{+,-\}^{\otimes L-1}}}\hspace{-.6cm} C_z(t) q^{n_-}, \label{eq:Z_weighted_sum}
\end{align}
where $n_-$ is the number of $-$ spins in $\ket{z}$. Importantly, only configurations that have a $\ket{+}$ at location $x$ can contribute to the OTOC. This is enforcing the operator insertion $W(x,0)$ at time $t=0$. While both $+$ and $-$ are allowed at other sites (since the effective spin basis is non-orthogonal), configurations that deviate from $\bra{-\cdots-+^* -\cdots -}$ are suppressed by powers of $q$. This leads naturally to the concept of a \textit{dominant state}: the configuration from $\{+,-\}^{\otimes L}$ that maximally contributes to $Z_{\rm OTOC}$ will have a $\ket{+}$ at position $x$ and $\ket{-}$ elsewhere.

{\bf Dominant state}: The {\it dominant state} is the configuration evolved through $\mathcal{M}(t)$ in the $\{+,-\}$ basis with the maximum contribution to $Z_{\text{OTOC}}$, i.e.,
\begin{equation}
    z_{\max}(t) =\argmax_{z \in \{+,-\}^{\otimes L}  } C_z(t) q^{n_-}.\label{eq:dominant_state}
\end{equation}
The bottom boundary condition in the OTOC \eqref{eq: otoc_Z} is critical in determining this state, as the bulk evolution itself cannot reveal the dominant state until the overlap between the bra state and $\ket{z(t)}$ is taken.
To find the dominant state at intermediate times, we introduce a modified dynamics that incorporates the $q^{n_-}$ weighting directly into the evolution. %
\begin{equation}
    M_{\rm modi} (a_x, a_y, a_z) = 
    \begin{pmatrix}
    1 & h/q  & h/q & 0\\
    0 & b_+ & b_- & 0\\
    0 & b_- & b_+ & 0\\
    0 & q h & q h & 1
    \end{pmatrix}.
\label{eq: modified transition mat}
\end{equation} 	
The modified transition matrix keeps track of the number of $-$ by increasing the weight by a factor of $q$ when the number of $-$ increases by one, i.e. when $\ket{+-}$ or $\ket{-+}$ transitions to $\ket{--}$ and a factor of $\frac{1}{q}$ when it decreases by one, i.e. when either of those states transitions to $\ket{++}$. Hence, using the modified dynamics, the dominant state can be equivalently computed as
\begin{equation}
   z_{\max}(t) = \argmax_{z^* \in \{+,-\}^{\otimes L}  } \langle z^* |  \prod_{t'=1}^{t} {\mathcal{M_{\rm modi}}(t')} | - +\cdots + \rangle,
\end{equation}
where $|z^* \rangle$ is in the dual basis of $|z\rangle$. 

$z_{\rm max}(t)$ is a snapshot of the largest weight at time $t$. It might be tempting to conclude that this state also dominates the sum before the final time slice. This is the case for the half-system purity, where $z_{\rm max}(t)$ at the first stage corresponds to a wandering domain wall, which is also the equilibrium configuration of the partition function. 

However, this picture does not hold for OTOC. The bottom boundary condition eliminates states with a $-$ at the location of $+^*$, while the intermediate states leading to $z_{\rm max}(t)$ are not subject to this constraint. %
In fact, starting from $|-+\cdots +\rangle$, a trajectory of states violating the constraint at intermediate stages can acquire a lower decay rate than simply connecting the dominant states at each time slice. %
This leads us to define a {\it dominant trajectory} below.

\begin{figure}
    \centering
    \includegraphics[width=0.9\linewidth]{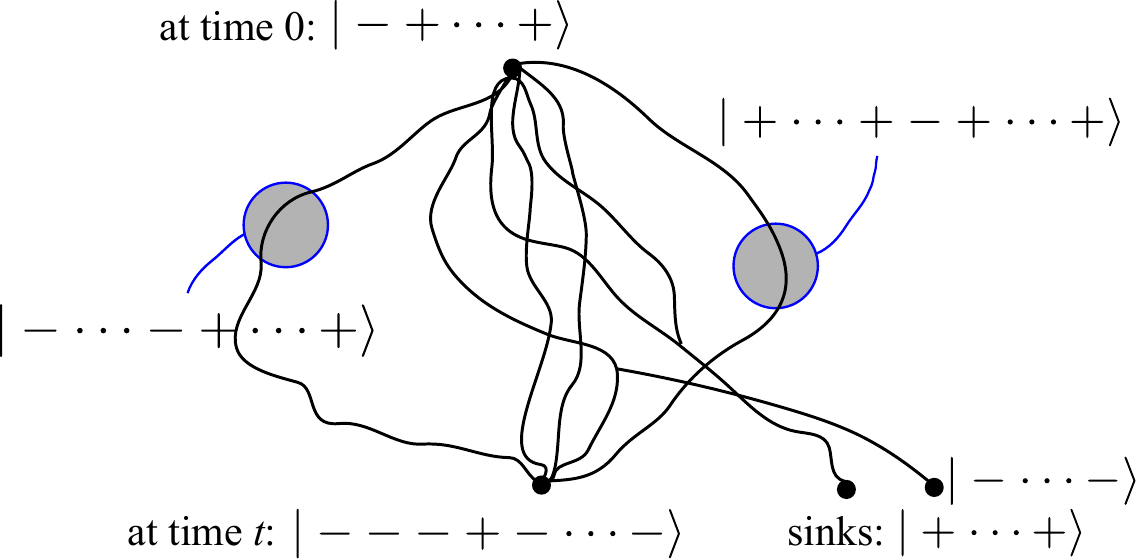}
    \caption{
    A schematic of different trajectories that give rise to the same dominant state at time $t$. The all $+$ and all $-$ states are sink states since they correspond to stationary points of the transition matrix \eqref{eq: transition mat}. The magnified trajectories show possible intermediate states (domain wall and a $-$ magnon) which would violate the boundary constraint if imposed at this intermediary time. However, both can evolve to the dominant magnon state at time $t$.
    }
    \label{fig: traj}
\end{figure}

{\bf Dominant trajectory}: The dominant trajectory is the most likely history that becomes the dominant state $z_{\rm max} (t)$ at time $t$. Mathematically, it can be probed by inserting a complete set of intermediate states and taking the largest state at time $\tau$ given the initial and the dominant state at the final time slice: %
\begin{align}
    \psi(\tau) &= \max_{z \in \{+,-\}^{\otimes L} } \bra{z^*_{\rm \max}} \prod_{t'=\tau+1}^t {\mathcal{M_{\rm modi}}}(t') \ket{z} \nonumber \\
    & \hspace{1.5cm} \times \bra{z^*} \prod_{t'=1}^\tau {\mathcal{M}_{\rm modi}(t')} \ket{ - + \cdots + }. \label{eq:dominant_trajectory}
\end{align}
In the picture of two-dimensional statistical mechanics, the dominant state is the bottom boundary condition, and the dominant trajectory $\{\psi(t')\}_{t'=1}^t$, is the 
equilibrium configuration. Fig.~\ref{fig: traj} shows multiple trajectories that satisfy the same boundary conditions but have different intermediate and sink states that 
only contribute to the time-independent saturation value. 

To visualize the dominant trajectory, we probe the magnetization at $(x, \tau)$
\begin{align}
    \langle \sigma_z(x, \tau)\rangle &:= \frac{1}{\mathcal{Z}(t)} \langle z_{\rm max}(t)^* | \prod_{t'=\tau+1}^t \mathcal{M}(t') \nonumber\\
    & \times \sigma_z(x,0) \prod_{t'=1}^\tau \mathcal{M}(t')  | - \cdots - + - \cdots -\rangle
    \label{eq: magnetization}
\end{align}
where $\sigma_z = | + \rangle \langle +^* | - | - \rangle \langle -^* |$ and $\mathcal{Z}(t)$ sums over all the possible trajectories leading to $z_{\rm max} (t)$. This computes the magnetization of the overlap between the boundary states evolved forward and backward by time $\tau$ and $t-\tau$ respectively. In this case,  we can use $\mathcal{M}$ instead of $\mathcal{M}_{\rm modi}$, since they only differ by a constant (in $t$) factor when the final state is fixed. The magnetization shows the dominant states while taking into account the degeneracies due to the $+ \leftrightarrow -$ symmetry of the bulk. The results are shown in Sec.~\ref{sec:averaged}.

\section{Results: averaged dynamics}\label{sec:averaged}

This section covers the numerical results for the random dual unitary ensemble $(a_x = 1, a_y = 1, a_z)$ defined below Eq.~\ref{eq:cartan-u}. All dynamics in this section are Markovian and in terms of the effective spin degrees of freedom. The results are summarized in Table~\ref{table:decay_rates}.

For the numerical computation, we perform a matrix-product state simulation of the Markovian dynamics of the random dual unitary circuits. 
Importantly, we invoke an orthonormal basis to extract the weights $C_z(t)$ in Eq.~\ref{eq:Z_weighted_sum}. The Markovian dynamics is linear and the inner product of the basis does not affect $C_z(t)$. Rather than using $(|+\rangle$, $|-\rangle)$ as a basis, we use an orthonormal basis $(|+), |-))$ 
, the computational basis, denoted by parenthesis instead of angle brackets: 
$\left\{ |+) = \begin{pmatrix}
    1\\0
\end{pmatrix}, |-) = \begin{pmatrix}
    0\\1
\end{pmatrix} \right\}$. This orthonormal basis differs from the effective basis $|+\rangle$, $|-\rangle$ used earlier. When computing \eqref{eq:dominant_state},\eqref{eq:dominant_trajectory}, \eqref{eq: magnetization}, we thus replace $\ket{z}$ with $|z)$ and $\bra z^*$ with $( z|$.

\begin{table}[H]
\centering
\begin{tabular}{|c|c|c|}
\hline
 & $r_1$  & $r_2$ \\
\hline
\textbf{BW}, OBC & $\frac{r_{\text{mag}}}{2}$ & 
\begin{tabular}{@{}c@{}}
$a_z <1/3$: $r_{\rm DW}$ \\
$a_z >1/3$: $r_{\text{mag}}$
\end{tabular} \\
\hline
\textbf{BW}, PBC & $\frac{r_{\text{mag}}}{2}$ & $r_{\text{mag}}$ \\
\hline
\textbf{S}, OBC & $\frac{r_{\text{mag}}}{2}$ & 
\begin{tabular}{@{}c@{}}
$a_z <1/3$: $r_{\rm DW}$ \\
$a_z >1/3$: $r_{\text{mag}}$
\end{tabular} \\
\hline
\textbf{S}, PBC & $\frac{r_{\text{mag}}}{2}$ & 
\begin{tabular}{@{}c@{}}
$\frac{r_{\text{mag}}}{2}$ \\ 
\end{tabular} \\
\hline
\end{tabular}
\caption{Decay rates for BW and S configurations under OBC and PBC conditions (a general extension of Tab.~\ref{tab:bz_otoc_summary}) Here, $r_{\text{DW}} = 1 $ and $r_{\text{mag}} = \ln \frac{3}{2- \cos(\pi a_z)}/\ln 2$. }
\label{table:decay_rates}
\end{table}

\subsection{The two-stage rates}
\label{subsec:the_2_stage_rates}
\begin{figure}[ht]
    \centering
    \begin{minipage}[t]{0.48\textwidth}
        \centering
        \includegraphics[width=\linewidth]{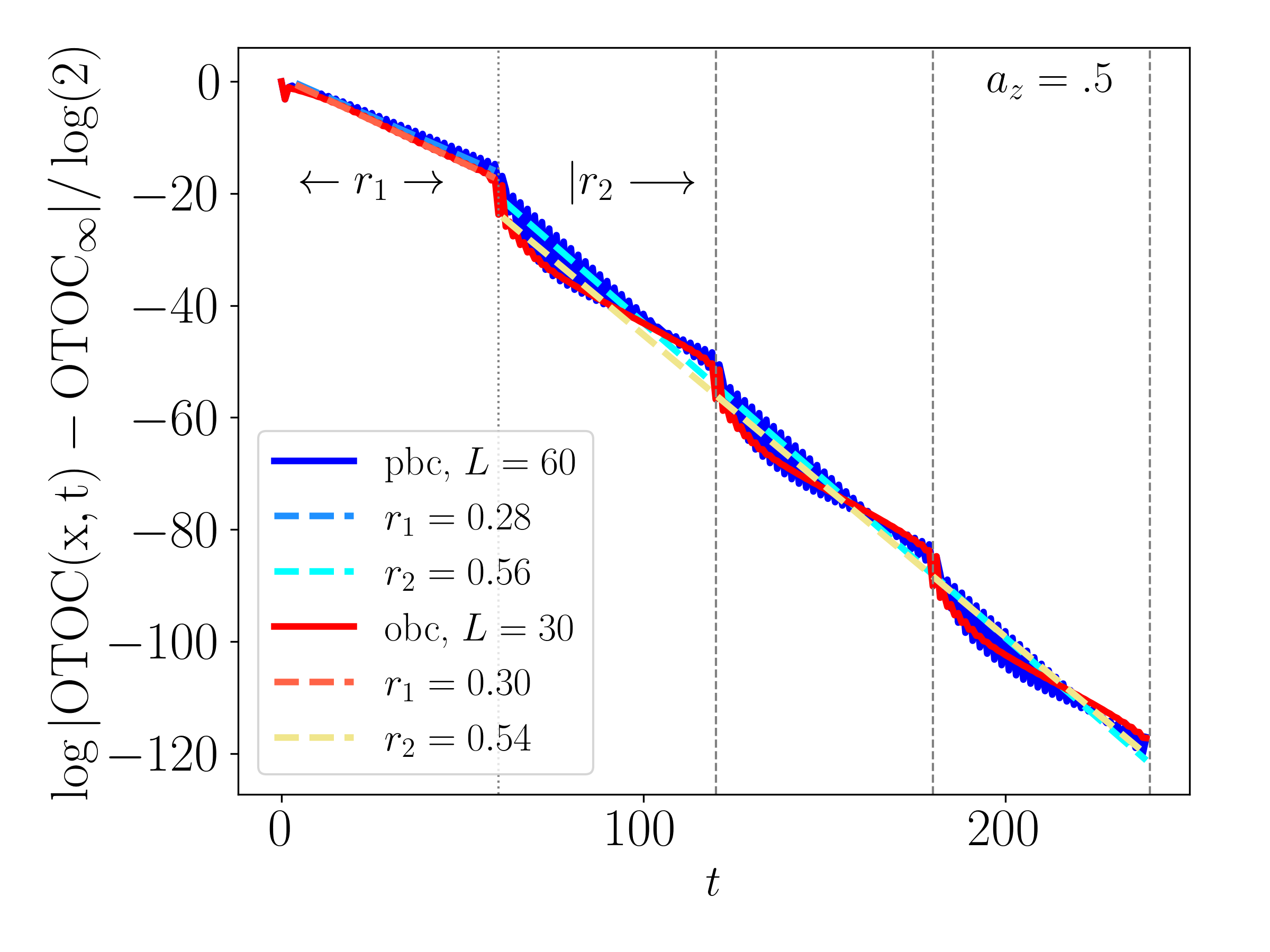}
        \caption{OTOC relaxation in brickwork geometry with $L=60$ (open BC, blue) and $L=30$ (periodic BC, red), setting $x=1$ in $W(x,0)$ and $a_z=0.5$. Vertical lines mark the $L$ multiples, showing boundary effects. OBC system halved for matching PBC periodicity.}
        \label{fig:BW_OTOC_convergence_pbc_vs_obc}
    \end{minipage}%
    \hfill
    \begin{minipage}[t]{0.48\textwidth}
        \centering
        \includegraphics[width=\linewidth]{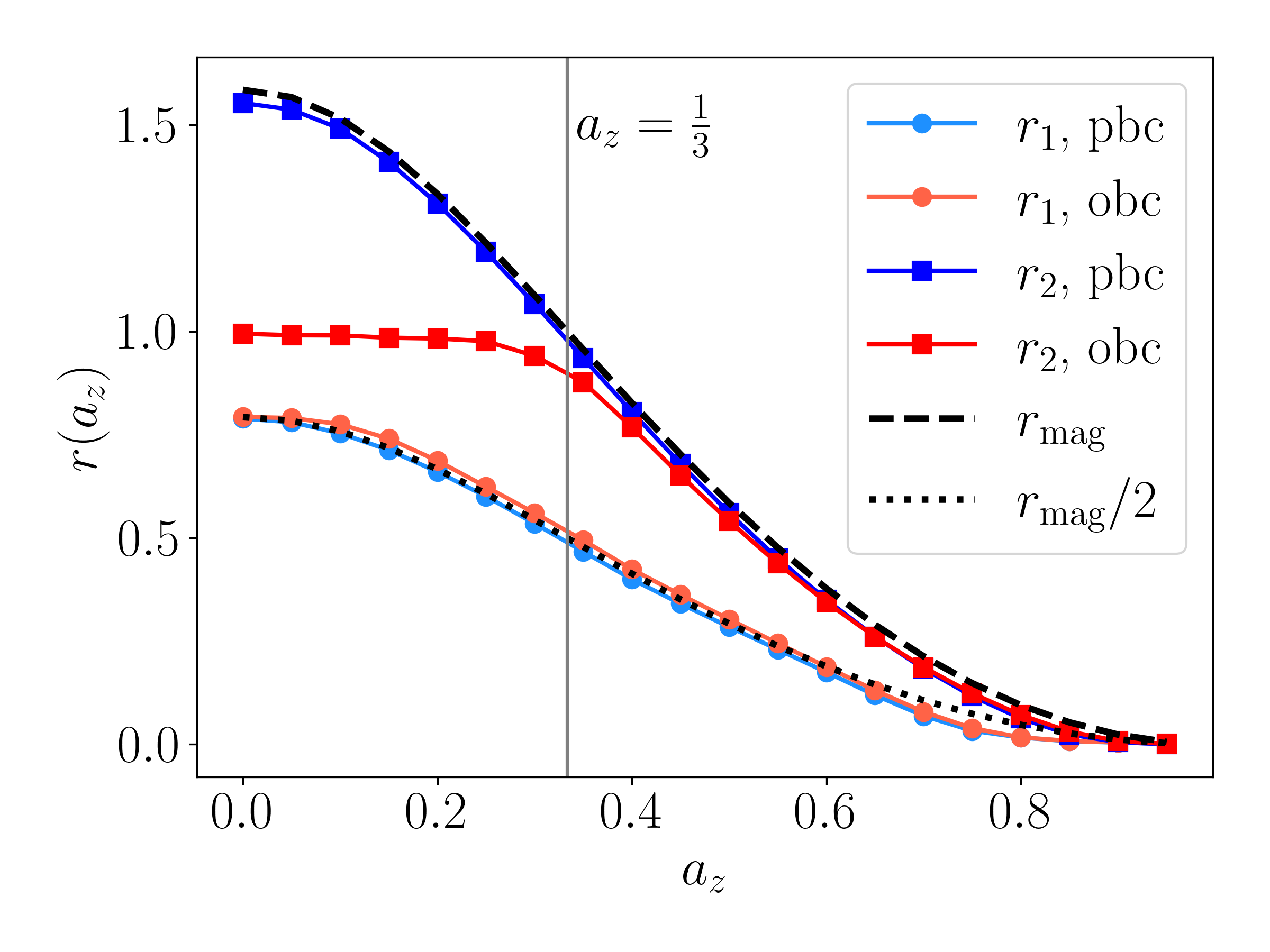}
        \caption{Decay rates $r_1(a_z)$ and $r_2(a_z)$ of OTOC relaxation in \bw circuit, obtained via MPS algorithm of effective Markovian dynamics. We use $L=60$ for $r_1$ fit and $L=20$ for $r_2$ extraction to ensure sufficient $L$ periods.}
        \label{fig:r1_r2_OTOC_BW}
    \end{minipage}
\end{figure}

Fig.~\ref{fig:BW_OTOC_convergence_pbc_vs_obc} illustrates the relaxation of the OTOC for both open boundary conditions (OBC) and periodic boundary conditions (PBC) in a brickwork (\bw) circuit with gates parametrized by $a_z=0.5$. The window of $r_1$ is $2L$ for OBC and $L$ for PBC. For visual comparison, we consider system sizes $L=30$ and $L=60$ respectively so that the time window of $r_1$ aligns. Vertical dashed lines indicate integer multiples of the system size $L$ and highlight discontinuities as the magnon bounces off the boundary (OBC) or wraps around (PBC). We use the same approach to find the rates in the range $a_z \in [0, 1]$. From this, we construct a phase diagram for $r_1$ and $r_2$ as a function of $a_z$, shown in Fig.~\ref{fig:r1_r2_OTOC_BW}. The first stage rate (circles) of both PBC and OBC are close to the rate $r_{\rm mag}/2$. The second stage rate (squares) is close to $r_{\rm mag}/2$ for PBC, and close to $r_{\rm mag}$ for OBC as long as $a_z<1/3$. This is to be contrasted with earlier works (see Tab.~\ref{tab:bz_otoc_summary}),%
which suggested that the second stage rate is $r_{\rm mag}$ irrespective of $a_z$. Instead, we find that for systems with open boundaries, values of $a_z < \frac{1}{3}$ cap $r_2$ to $1,$ %
a pattern identical to the second stage decay of the purity \cite{jonay_physical_2024}.

\begin{figure}[ht]
    \centering
    \begin{minipage}[t]{0.48\textwidth}
        \centering
        \includegraphics[width=\linewidth]{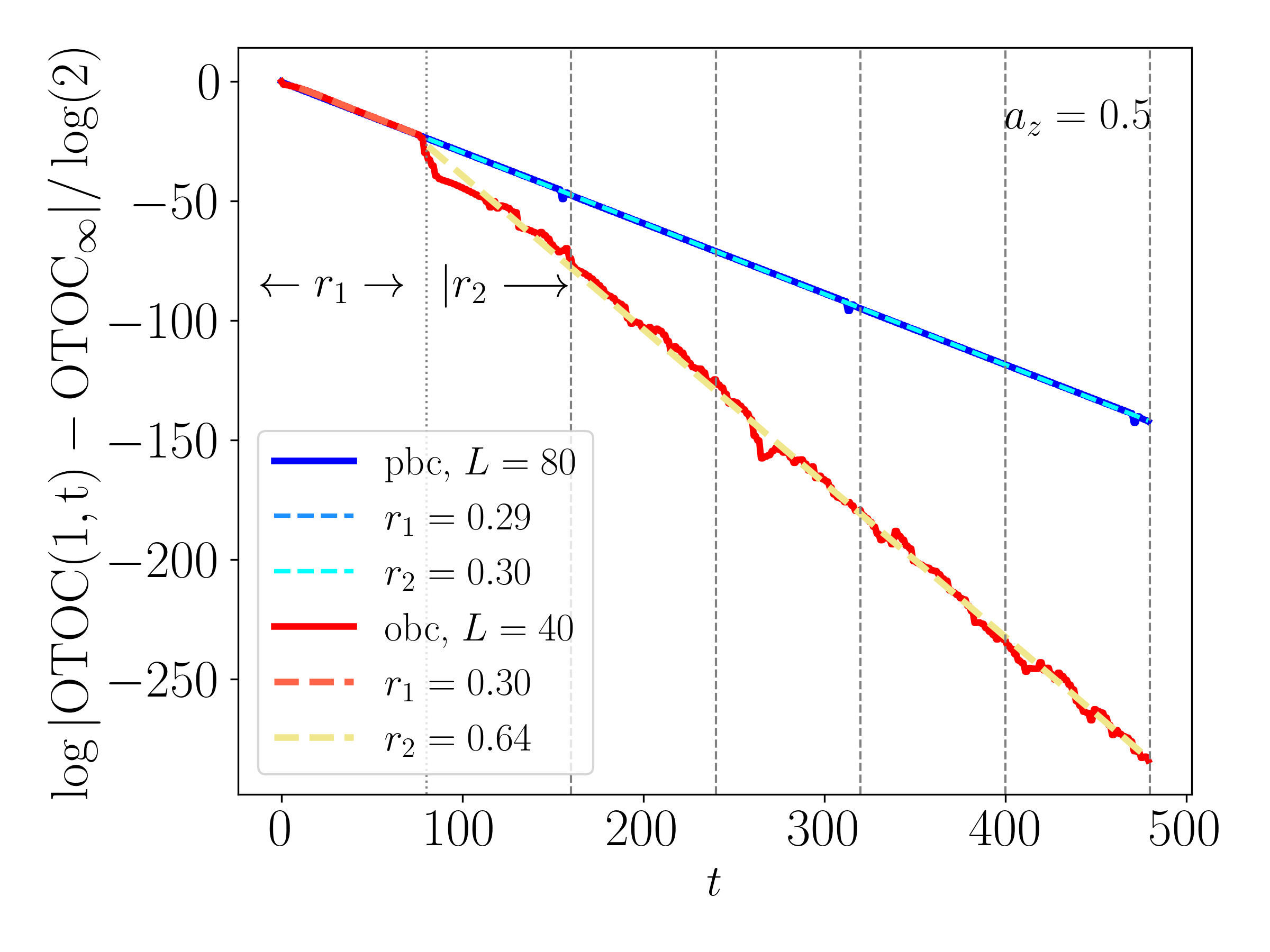}
        \caption{Relaxiation of the OTOC in \stair geometry: We take $L = 60$ for OBC (blue) and $L = 30$ for PBC (red). We set $a_z=0.5$.%
        }\label{fig:S_OTOC_convergence_pbc_vs_obc}
    \end{minipage}%
    \hfill
    \begin{minipage}[t]{0.48\textwidth}
        \centering
        \includegraphics[width=\linewidth]{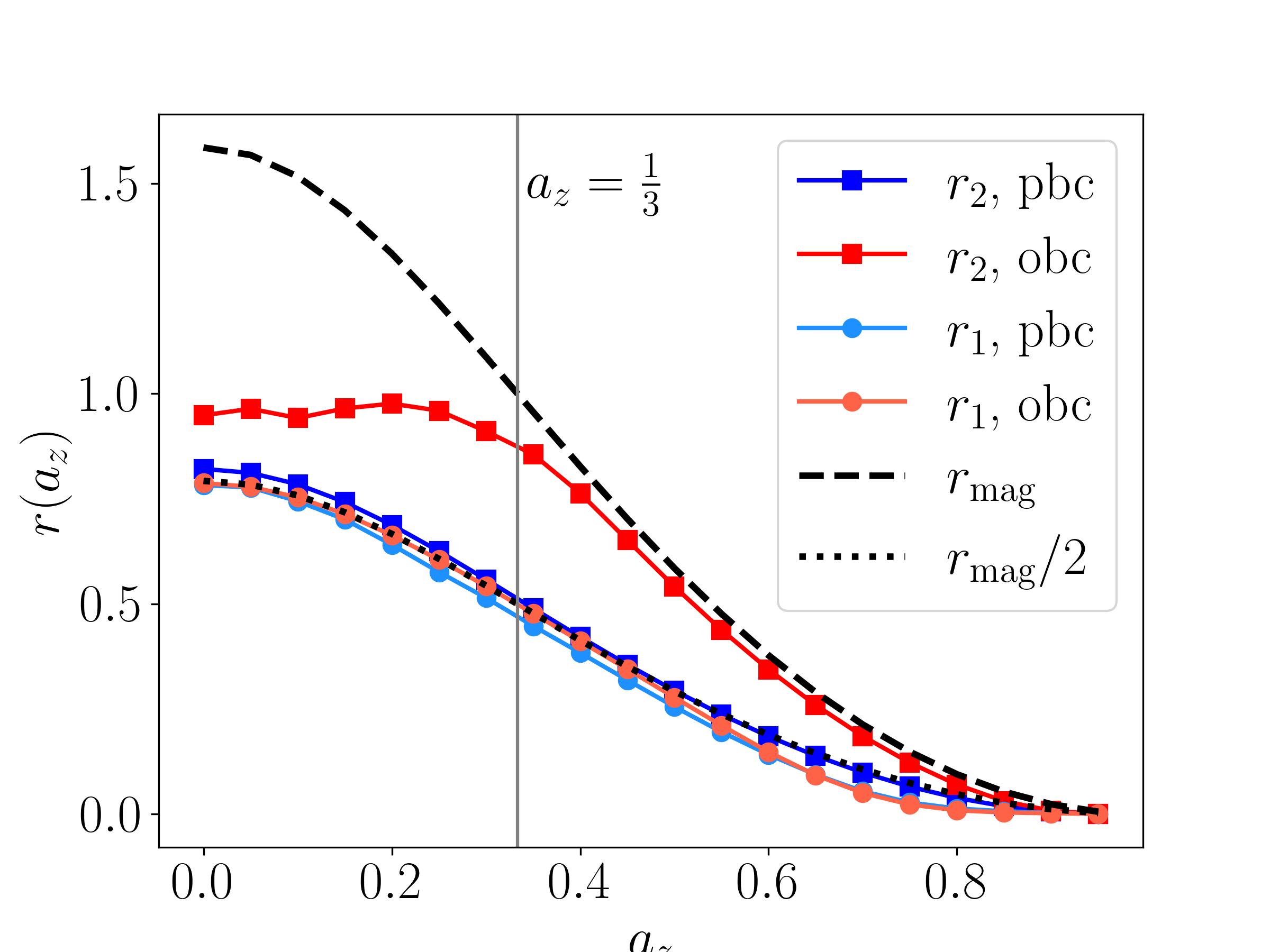}
        \caption{Decay rates $r_1(a_z)$ (first stage) and $r_2(a_z)$ (second stage) of the OTOC relaxation to its saturation value in the \stair geometry.}
        \label{fig:r1_r2_OTOC_S}
    \end{minipage}
\end{figure}

In Figs.~\ref{fig:S_OTOC_convergence_pbc_vs_obc} and Fig.~\ref{fig:r1_r2_OTOC_S}, we repeat the analysis above for the \stair geometry. The results are almost identical to the \bw geometry, except for the second stage of PBC, which produces %
$r_{\rm mag}/2$ instead of $r_{\rm mag}$. A detailed discussion of the physical mechanisms underlying these observations is presented in Sec.~\ref{subsec:dom_traj} below.

\subsection{Further numerical implementation details}\label{subsec:mps_num_details}

We use a standard matrix product state (MPS) to represent $\prod_{\tau =1}^t \mathcal{M}(\tau)| - +\cdots +)$. 
The relaxation process we are interested in has an exponential decay in amplitude. Consequently, our observables are exponentially small numbers at late times. We implement the following numerical procedures to improve precision. 

First, we remove the weights of the two equilibrium states $|+\cdots +)$ and $|-\cdots -)$ at each time step. These states are sinks of the Markovian dynamics
They are static under the evolution of any transition matrix resulting from unitaries and only contribute to the saturation value. In our partition function, the saturation value of the Markovian dynamics is of order one. When combined with the exponentially small quantity of interest, the arithmetic errors due to machine precision of double floating numbers can shadow the exponentially small observables. By subtracting their weights, we not only remove the infinite-time value from the OTOC at each step of the simulation but also use the double floating precision number to only represent the exponentially small quantity, which helps the numerical precision. 

After removing the weights of the sink states, all the other weights in $\sum_{\tau = 1}^t \mathcal{M}(\tau) |- + \cdots + )$ decay exponentially. Because of the non-unitary stochastic dynamics, %
the state vector is not normalized. To define a quantum mechanical analog of the wave function, we thus compute $N(t) | \sum_{\tau=1}^{t} \mathcal{M}(\tau) | - + \cdots + )$. Importantly, we store the normalization factor $\log(N(t))$ at each time point separately from the weights. %
To compute the OTOC at the end of the computation, we restore the factor of $\exp( - \sum_{\tau=1}^t N(\tau) )$. It avoids underflow or overflow errors from multiplying numbers of very different scales.

Despite these attempts, the MPS approximation error is still subject to the machine precision of floating numbers. Beyond system size $L = 60$, the signal becomes too weak at late times, even though runtime would allow us to simulate much larger systems. This limitation cannot be remedied by increasing the bond dimension, it is caused by precision issues rather than lack of computational resources.

{\bf Periodic boundary condition}: As a note, the matrix product state simulation of the periodic boundary condition is significantly more expensive than the open boundary. Connecting the last and first lattice doubles the entanglement. For the simulation, we have to swap the rightmost qubit across the lattice, act with the boundary gate, and swap it back. The system sizes in this paper were reasonably small so that we can simulate the periodic boundary. An alternative trick for the first stage would be to place an operator in the middle of the chain and use an open boundary circuit. 

\subsection{Dominant trajectories}
\label{subsec:dom_traj}
We now analyze the average magnetization profile and extract the decay rates shown in Tab.~\ref{table:decay_rates} from the dominant trajectories. In our diagrammatic convention, time flows downward with the initial state at the top.

\begin{figure}
    \centering
    \includegraphics[width=1 \linewidth]{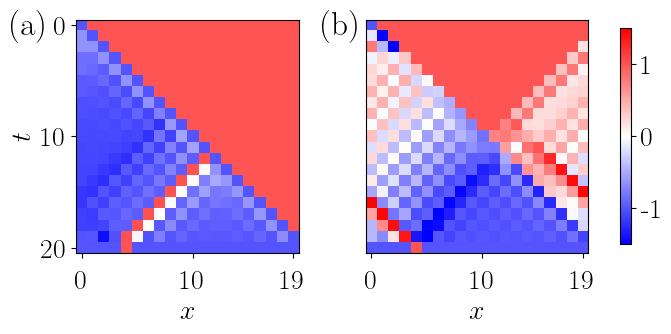}
    \caption{The short-time ($t \in (x, L]$) averaged magnetization corresponding to the OTOC evolved through the \bw geometry for (a) open boundary condition (OBC); (b) periodic boundary condition (PBC).
    We use $a_z = 0.2$ and $L = 20$. The $-$ spin on the top is placed at $x=0$ and the $+$ is placed at $x = 4$ at the bottom.}
    \label{fig: BW stage 1}
\end{figure}

{\bf Brickwork geometry}: Fig.~\ref{fig: BW stage 1} shows the first stage decay of the OTOC in a BW geometry. We place $V$ at $x = 0$ and $W$ at $x = 4$. Due to the weight imposed by the final state, the dominant state $z_{\rm max}(t) $ is identical or close to a $+$ magnon at $x = 4$. %
The magnetization heatmap reveals the dominant trajectory. For OBC, the initial $-$ magnon propagates as a domain wall on the light cone to create a large domain of $-$ that is favored by the bottom boundary condition. To satisfy the requirement of having an $+$ at $x_f$, it splits out a $+$ magnon with the shortest possible length aiming towards $x_f$. 

\begin{figure}[ht]
    \centering
    \includegraphics[width=.8 \linewidth]{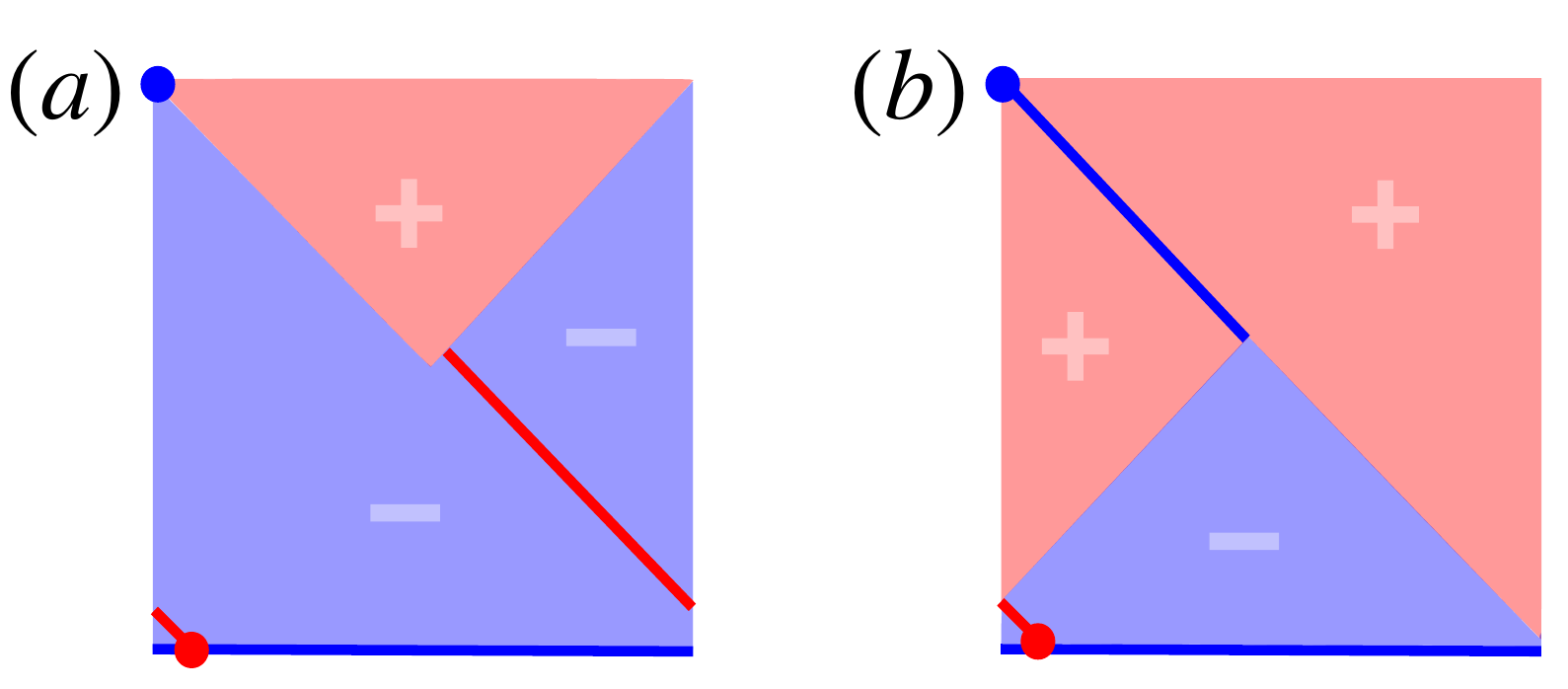}
    \caption{%
    Schematics for the degenerate trajectories for the PBC at short times. Superimposing them results in the magnetization heat map in the right panel of Fig.~\ref{fig: BW stage 1}. }
    \label{fig: short time deg}
\end{figure}

The time dependence of the weight can be computed as follows: The domain wall cost is approximately $\exp( - r_{\rm DW} t \ln q ) = q^{-t}$; however by creating a $-$ domain of size $t$, the bottom boundary will reward a factor of $q^{t}$, exactly canceling the domain wall cost. What remains is the cost of a magnon, which is $\exp(- \ln (q)  r_{\rm mag} (t-x)/2)$. Thus the decay rate is $r_{\rm mag} / 2$. 

We take $x$ slightly larger than $0$ to break the left-right reflection symmetry of the system. In the OBC, this displacement eliminates the degeneracy completely, so that the magnetization looks very similar to the unique dominant trajectory. 

The calculation for the PBC is similar. The difference is that there are now two domain walls at the initial stage. They can close and annihilate in $L/2$ instead of $L$ steps, resulting in half the transition time. There are degeneracies in the dominant trajectories, as shown in Fig.~\ref{fig: short time deg}. The superposition of the degenerate trajectories creates regions with close-to-zero magnetization shown in Fig.~\ref{fig: BW stage 1} (b).

\begin{figure}
    \centering
    \includegraphics[width= \linewidth]{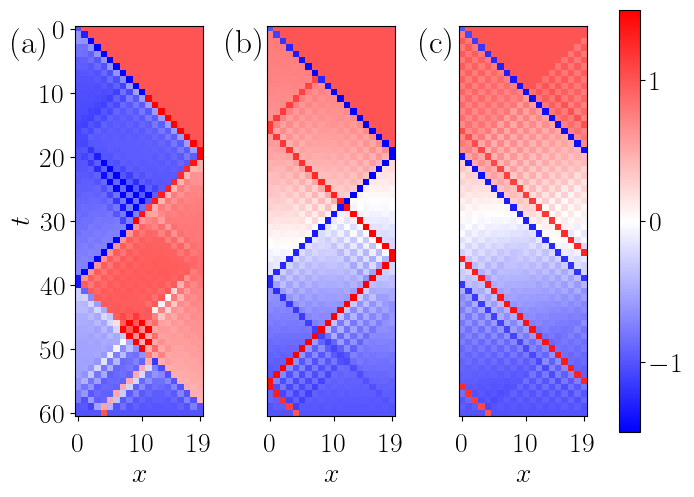}
    \caption{The long-time ($t\sim 3L$) magnetization corresponding to the OTOC evolved through \bw geometry with $L = 20$ for (a) $a_z = 0.2$, OBC, (b) $a_z = 0.7$, OBC, and (c) $a_z = 0.7$, PBC. The operator on the top is placed at $x=0$ and the operator at the bottom is placed at $x = 4$ .}
    \label{fig: BW stage 2}
\end{figure}

\begin{figure}
    \begin{subfigure}[t]{.1\textwidth}
        \includegraphics[height=5cm]{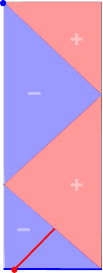}\\
        \vspace{.2 cm}
        \includegraphics[height=5cm]{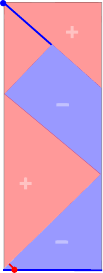}
        \captionsetup{justification=centering}
        \caption{DW degeneracies }
        \label{fig: dw_deg_obc}
    \end{subfigure}
    \hspace{.4cm}
    \vrule
    \hspace{.4cm}
    \begin{subfigure}[t]{.1\textwidth}
        \includegraphics[height=5cm]{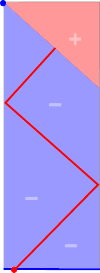}\\
        \vspace{.2 cm}
        \includegraphics[height=5cm]{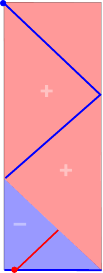}
        \captionsetup{justification=centering}
        \caption{Magnon degeneracies for OBC}        
        \label{fig: mag_deg_obc}
    \end{subfigure}
    \hspace{.4cm}
    \vrule
    \hspace{.4cm}
    \begin{subfigure}[t]{.1\textwidth}
        \includegraphics[height=5cm]{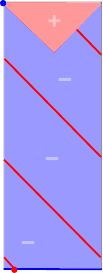}\\
        \vspace{.2 cm}
        \includegraphics[height=5cm]{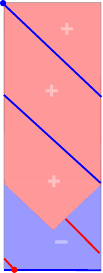}
        \captionsetup{justification=centering}
        \caption{Magnon degeneracies for PBC}
        \label{fig: mag_deg_pbc}
    \end{subfigure}
    \caption{Examples of possible degenerate trajectories in the long time regions for the relaxation of the OTOC. Red indicates $+$ spins and blue indicates $-$. %
    }  
\end{figure}

In the long time limit, the dominant state at the final time slice is always close to a magnon, but the intermediate state can be a domain wall. It incites a competition. Fig.~\ref{fig: dw_deg_obc} shows  two long-time trajectories whose bulk part can be either domain wall or magnon. Aside from the costs in regions close to the top and bottom boundaries, the OBC partition function is $\exp( - r_{\rm DW} t \ln q )$ or $\exp( - r_{\rm mag} t \ln q )$. When $a_z < \frac{1}{3}$, the domain wall rate is smaller and $r_2 = r_{\rm DW} = 1$. Otherwise, $r_2 = r_{\rm mag}$. Fig.~\ref{fig: mag_deg_obc} and \ref{fig: mag_deg_pbc} show the degenerate trajectories in which the time a magnon shows up can be different due to $+/-$ exchange symmetry. The averaged magnetization $\langle \sigma^z \rangle$ effectively computes the superpositions of all possible magnon trajectories.

{\bf Staircase geometry}: As mentioned in Sec.~\ref{sec:review}, the staircase structure can be implemented by applying the gates sequentially in each layer, which goes upwards in the light cone direction. Each layer corresponds to two layers in the \bw circuit when it comes to counting time as there are twice as many gates. The PBC is implemented such that the two-qubit gate is applied on the first and last sites. In a spacetime diagram, these two sites are connected by a segment of roughly $45^\circ$ angle in the light cone direction. This creates a shortcut because the first and last sites are adjacent in gates but also $L-1$ apart in both the vertical and the horizontal directions in spacetime.

As demonstrated in Fig.~\ref{fig:bw_s}(c), the staircase circuit gives the same OTOC decay rate as that of the \bw circuit for the OBC. 
Pictorially, the domain wall is still of the length $t$ and cancels with the boundary contribution $2^{L-t}$; the magnon is of length roughly $t/2$. These combinations of factors give $\frac{r_{\text{mag}}}{2}$ as the decay rate. 

\begin{figure}
    \centering
    \includegraphics[width=.8\linewidth]{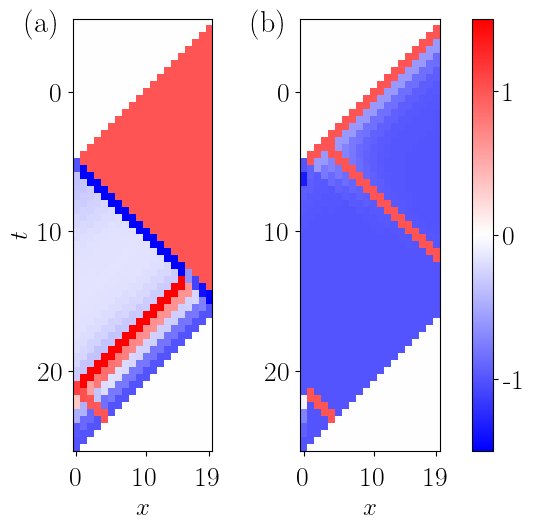}
    \caption{Magnetization for the S geometry at $a_z = 0.2$ for (a) OBC and (b) PBC. The system size is $L=20$.}
    \label{fig:S_heatmap}
\end{figure}

\begin{figure}
    \includegraphics[width=.7\linewidth]{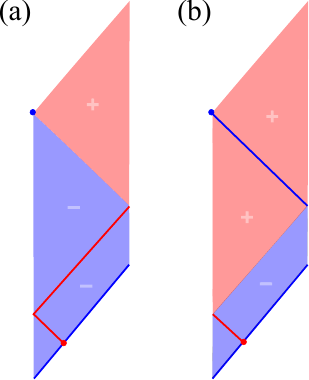}
    \caption{The two possible degenerate dominant trajectories in the \stair geometry for OBC. Superimposing them results in the magnetization heatmap in Fig.~\ref{fig:S_heatmap} (a).} 
\end{figure}

\begin{figure}
    \centering
    \includegraphics[width=1\linewidth]{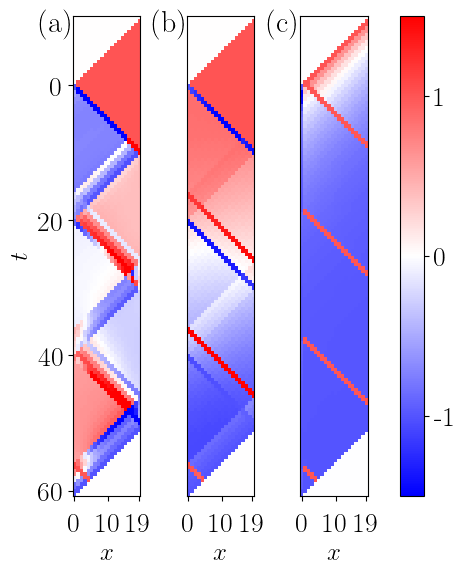}
    \caption{The long time ($t = 4.5L$) magnetization for the \stair geometry with $L = 20$ for (a) $a_z = 0.2$, OBC; (b) $a_z = 0.7$, OBC; (c) $a_z = 0.2$, PBC. The operator on the top is placed at $x=0$ and the operator at the bottom is placed at $x = 4$.}
    \label{fig: S stage 2 }
\end{figure}

At long times, the dominant modes are analogous to that of the BW circuit, where the dominant trajectories depend on $a_z$ in case of OBC, while the PBC geometry favors the the magnon trajectory ubiquitously. The only difference in the PBC case is that the magnon trajectory is reduced by half, due to the $45^\circ$ tilt of the geometry so that the magnon can skip $L$ steps from the other side of the boundary, as shown in Fig. ~\ref{fig: S stage 2 } (c).

Overall, the boundary condition can only contribute a difference of $q^L$  for each configuration, which pales in comparison with the exponential decay in $t$ when time is large. Therefore the long-term decay rate will always be determined by the bulk modes, namely the result of the competition between the domain wall and magnon.

\section{Results: clean, time translation invariant systems} \label{sec:floq}

\begin{figure}[!h]
    \centering
    \includegraphics[width=1\linewidth]{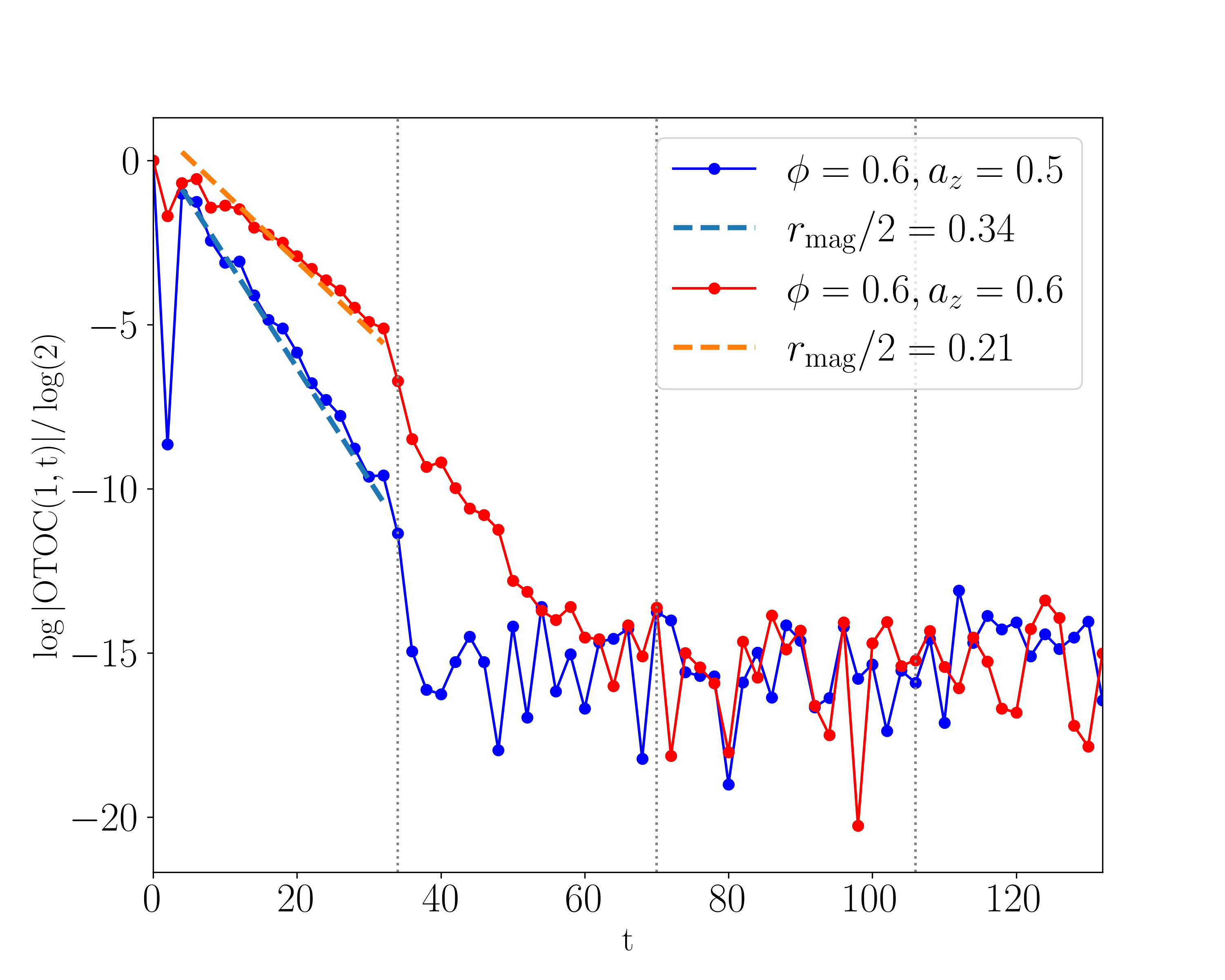}
    \caption{OTOC relaxation under unitary Floquet dynamics (Eq. \eqref{eq:UFloquetZnidaric}) with $L=18$, open boundaries, and $\phi=0.6$. Initial decay follows predicted magnon rates. For $a_z=0.6$ (red), the dynamics exhibit a possible second relaxation stage, while for $a_z=0.5$ (blue), the system appears to thermalize directly after the initial magnon-dominated decay.
    }
    \label{fig:FloquetOtoc_fixedSystem}
\end{figure}

\begin{figure}
    \centering
    \includegraphics[width=1\linewidth]{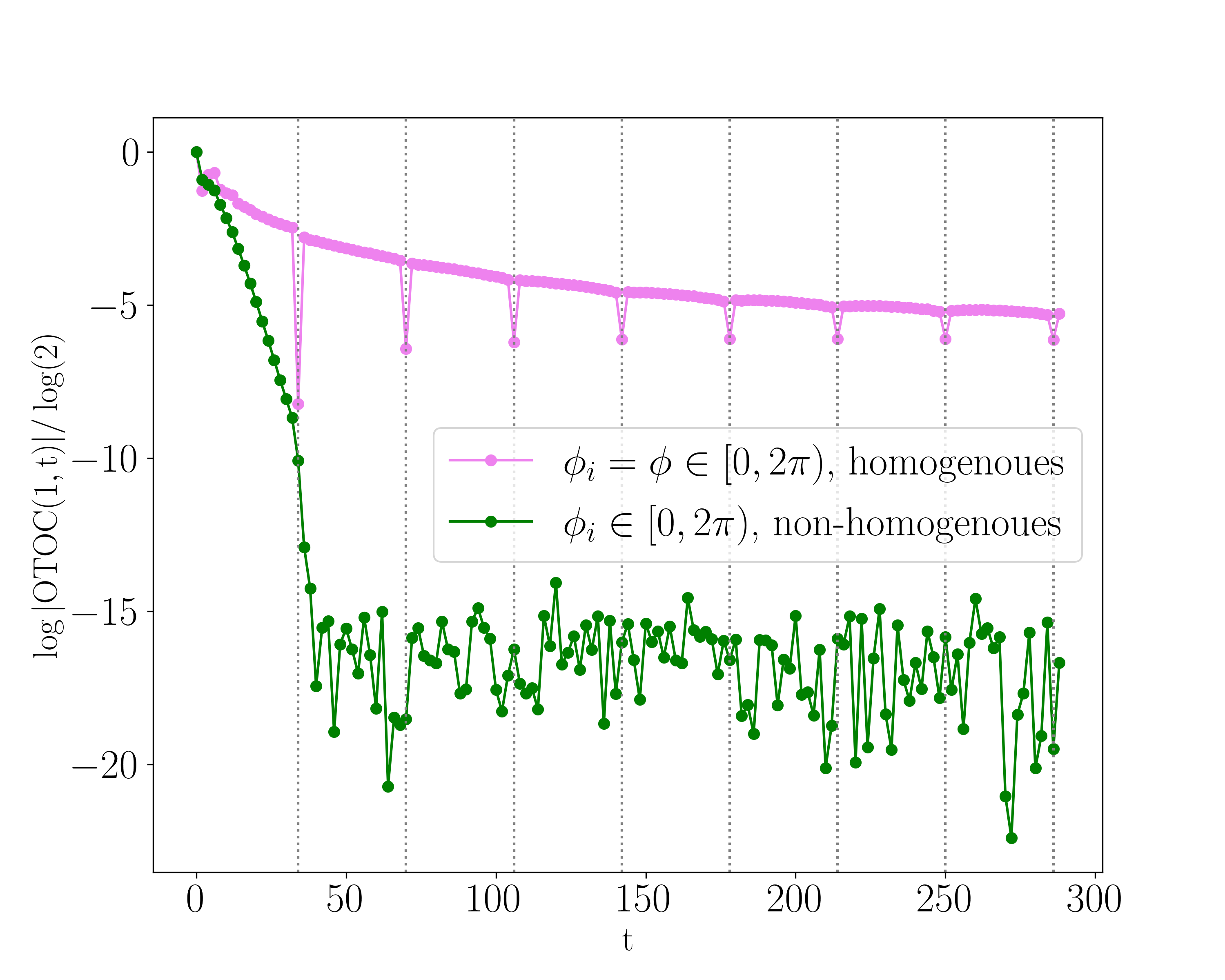}
    \caption{OTOC relaxation under averaged Floquet dynamics (Eq. \eqref{eq:UFloquetZnidaric}) with $L=18$, OBC and $a_z=0.5$. $\phi$ is sampled uniformly from $[0,2\pi)$. The homogeneous case (pink) in which $\phi$ is the same on all sites shows remarkably slow relaxation with periodic discontinuities due to the boundary effect. In contrast, the non-homogeneous case (green, independent $\phi_i$) thermalizes rapidly. The Haar-random saturation value is $-1/(4^L-1)$\cite{znidaric_two-step_2023}, approximately $O(10^{-36})$ for $L=18$.
    }
    \label{fig:FloquetOtoc_sampledSystem}
\end{figure}

In this section, we analyze the dynamics of clean and disordered Floquet systems. Recent work\cite{znidaric_two-step_2023} has demonstrated that the two-stage decay of out-of-time-ordered correlators (OTOC) persists without random averaging.%
Similarly, we hypothesize that our theory of emergent statistical model and the domain wall/magnon framework continues to predict the relaxation rates of both stages, albeit with renormalized free energies in the clean limit.

For clean systems, the effective degrees of freedom are no longer restricted to two spins since we are not averaging over disorder. Therefore, we compute the OTOC directly. Instead of evaluating the full trace ${\rm Tr}(\mathcal{O}(t))=\sum_{n=1}^{2^L} \langle n|\mathcal{O}(t)|n\rangle$, we use canonical typicality which approximates the exact trace by approximating it by an average over Haar random states: ${\rm Tr}(\mathcal{O}(t)) \approx \frac{1}{N} \sum_{i=1}^{N} \langle \psi_{i} | \mathcal{O}(t) | \psi_{i} \rangle$. For circuit models, this has the benefit that we never need to encode the full matrix $\mathcal{O}(t) = U^{\dagger}\mathcal{O}U$, but can calculate the quantity by acting locally with the constituent gates of $U$ on the vector $\ket{\psi_i}$. While we always choose $N \ll 2^L$ to make the approximation useful, for large system sizes $L$ a single random state $N=1$ is often sufficient for accurate results.

We consider a translation-invariant Floquet brickwork model characterized by a local 2-site unitary uniquely parametrized by $(\phi,a_z)$:
\begin{align}
     W(a_z)&= e^{-i \frac{\pi}{4} (X \otimes X +Y\otimes Y +a_z Z\otimes Z)}  u(\phi) \otimes u(\phi) \notag\\
     u(\phi)&=e^{i(\sin(\phi)X+\cos(\phi)Z)}. \label{eq:UFloquetZnidaric}
\end{align}
Note that the two-site gate is identical to the one in \eqref{eq:u_sym} with $a_x=a_y=1$ (the dual unitary case). We consider three important cases:

\textbf{Case 1} [Clean Homogenous Floquet]
In the clean homogeneous case $\phi=0.6$, we successfully reproduce the first-stage decay rates $r_1$, which align with our stochastic model predictions. The second-stage decay rate $r_2$ is only observable when $a_z \geq 0.6$. For systems with $a_z \lesssim 0.6$, complete thermalization occurs after the first stage. We have not yet identified any values of $\phi$ that produce second-stage decay for $a_z \lesssim 0.6$ in clean systems. This was the motivation to test thermalization in disordered systems.

\textbf{Case 2} [Disorder Homogenous Floquet]
In this setting, we sample $\phi$ uniformly in $[0,2\pi)$, but each circuit remains translation invariant. The OTOC in Fig.~\ref{fig:FloquetOtoc_sampledSystem} is the result of averaging over circuits with different values of $\phi$. We observe a strikingly slow decay in the second stage extending over multiple periods, with characteristics similar to a power-law decay. We have identified that this slow decay is caused by critical values of $\phi$ close to $\pi/2$. However, $\phi$ itself does not affect the entangling power of the gates in \eqref{eq:UFloquetZnidaric}. The precise conditions that give rise to this power-law-like behavior are under investigation.

\textbf{Case 3} [Disorder nonhomogeneous Floquet]
The disordered nonhomogeneous case demonstrates clear two-stage thermalization as predicted by our theoretical framework for random in time dynamics. This supports the assumption that it carries over to Floquet systems.

\section{The Cluster Picture}
\label{sec:cluster}

In addition to the spin basis, the cluster basis may also intuit operator spreading \cite{nahum_real-time_2022-1}. The cluster refers to the contiguous spatial regions occupied by traceless operators and\cite{nahum2018operator,chen_operator_2018,nahum_real-time_2022-1,qi_quantum_2018,vonKeyserlingk2018operator} the growth of the cluster size effectively describes operator spreading before $t = x / v_B$. In this section, we demonstrate that the magnon in the spin basis is equivalent to the bound state in the cluster basis, i.e., the operator remains local in space. The decay rate of relevant observables, e.g., the one-point function squared, can be directly read off from a matrix element of the two-site transfer matrix. We review the cluster basis, compute the convergence rates of the one-point function squared, and generalize the computation to OTOC. 

\subsection{Review of the cluster basis}
To compute the OTOC in the cluster basis, we first expand the operator at each time step as a superposition of the generalized Pauli strings, which is a tensor product of the Pauli basis $\{\oS\} = \{|\sigma^{ \mu }\rangle \}_{\mu = 0}^{q^2-1}$: 
\begin{equation}
O(x, t) = \sum_{\mathcal{S}} a_{\mathcal{S}} (t) \mathcal{S}. 
\end{equation} 

The coefficients $a_{\oS}(t)$ keep track of the amplitude of the weight of $\oS$ in the operator at time $t$. It can be calculated through the inner product\footnote{The normalization of the operators is $\Tr( \mathcal{S} \mathcal{S}' )  = \delta_{ \mathcal{S}  \mathcal{S'} } \Tr(\I)$. }
\begin{equation}
\label{eq: one-pt fcn coeffs}
a_{\oS} (t+1) = \frac{1}{\Tr(\I)} \Tr( U^\dagger O(x,t) U \mathcal{S}^\dagger ).
\end{equation}
Its evolution can be streamlined by a $q^{2L} \times q^{2L}$ transfer matrix $\mathbf{V}$: 
\begin{equation}
    a_{\oS}(t+1) = \sum_{ \oS'} \mathbf{V}_{\oS \oS'} a_{\oS'} (t) .
\end{equation}
In principle, $a_{\oS}(t)$ can evolve through a different trajectory from $a^*_{\oS}(t)$. However, quantum chaos dephases the different operator contributions to $|a_{\oS}(t)|^2$, and the off-diagonal terms are strongly suppressed. 
After random averaging, the cancellation is exact, leading to a Markovian dynamics for $|a_{\mathcal{S}}(t)|^2$:
\begin{equation}
|a_{\mathcal{S}}(t+1)|^2 = \sum_{ \mathcal{S}'} T_{ \mathcal{S} \mathcal{S}' } | a_{\mathcal{S}'}(t)|^2 .
\end{equation}

As before, we average the operator on each site over a single-qubit Haar unitary to obtain the transfer matrix on a cluster basis. Consequently, the effective space of $\oS$ is reduced to a $2^L$ state space, i.e., a site is either occupied by any traceless operator, denoted by $\fc$, or unoccupied, denoted by $\ec$. A cluster is any contiguous region of $\fc$. For example, the initial local traceless operator used to compute the OTOC is a cluster of size 1. Under unitary evolution, the cluster size will grow, typically with velocity $v_B$. 

The two states can be expressed in the operator space of the doubled Hilbert space as
\begin{equation}
| \ec \rangle  = \frac{\I\otimes \I }{q}  \quad \ket{\fc} = \frac{1}{q^2 - 1} \sum_{i=1}^{q^2 -1} | \frac{1}{q}\sigma^i \otimes \sigma^i  \rangle,
\end{equation}
with normalization: 
\begin{equation}
\label{eq: cluster_normalization}
\langle \ec | \ec \rangle  =  1, \quad \langle \fc | \fc \rangle  = \frac{1}{q^2 - 1} \text{ and } \braket{\fc}{\ec} =0.
\end{equation}

The averaged two-site transfer matrix in this basis is
\begin{equation}
\label{eq: transfer mat cluster}
T(1,1, a_z)
=
\begin{bmatrix}
1 & 0& 0&0 \\
0 & 0& \frac{2-\cos \pi a_z }{3}& \frac{1+\cos \pi a_z }{9}\\
0 & \frac{2-\cos \pi a_z }{3} & 0& \frac{1+\cos \pi a_z }{9}\\
0 & \frac{1+\cos \pi a_z}{3}& \frac{1+\cos \pi a_z}{3} & \frac{7-2\cos \pi a_z }{9}\\
\end{bmatrix}.
\end{equation}
This is a stochastic matrix due to the probabilistic interpretation of the Pauli string weights. Such transition matrix has been classified in Ref.~\cite{Akhtar_Anand_Marshall_You_2024}, and the constraints lead to only two free parameters.  Note that the exchange process $\ec \fc \leftrightarrow \fc \ec$ has a rate of $\frac{2-\cos \pi a_z }{3}$, precisely the magnon decay in one time step.

\subsection{One-point function squared}

In this section, we derive emergent dynamics of  the one-point function $\overline{|\bra{\psi} O(t)\ket{\psi}|^2}$, where $\psi$ is a tensor-product state, and show that a transition in the decay rates exists depending on the magnitude of $a_z$. In the spin basis, it is a binding transition of the two domain walls\cite{jonay_physical_2024}. The binding transition also occurs in the cluster basis (also see the random walkers introduced in \ref{subsec: mode}) 

As before, a local operator $O(0)$ is inserted at the initial time slice, which is the top boundary of the classical system. Since this is represented as $|\fc \ec \cdots \ec \rangle$ in the cluster basis, it is qualitatively the same as the top boundary condition in the spin basis $|-^* + \cdots + \rangle$ when computing the OTOC. In contrast, the bottom boundary condition differs, as there is no operator insertion, and $\psi$ is fixed as a tensor-product state. In the doubled Hilbert space representation, the bottom boundary is represented by the replicated state $|\psi \rangle  \otimes |\psi \rangle^* \otimes |\psi \rangle  \otimes |\psi \rangle^* \equiv |\psi^{\otimes 4} \rangle$. Hence, the one-point function squared in the cluster basis is
\begin{equation} 
\label{eq: one-pt fun 2}
\begin{aligned}
  &\overline{|\bra{\psi} O(t) \ket{\psi}|^2}  = \langle  \psi^{\otimes 4} |\prod_{\tau = 1}^t\mathcal{T}(\tau) \ket{\fc \ec \cdots \ec} \tr(O^2) q^{L-1}\\
  &= \sum_{p \in \{\ec, \fc \}^{\otimes L} } \langle  \psi^{\otimes 4}  | p \rangle  \langle  p^* |  \prod_{\tau = 1}^t\mathcal{T}(\tau) \ket{\fc \ec \cdots \ec}  \tr(O^2)q^{L-1}
\end{aligned}
\end{equation}
where $\mathcal{T}(t)$ is the transfer matrix constructed from tensor products of \eqref{eq: transfer mat cluster} at $t$. The final state contribution to $\overline{|\bra{\psi} O(t) \ket{\psi}|^2}$ can be obtained by inserting a complete set of basis $|p \rangle \langle  p^*|$, where $p$ stands for the $\ec \fc$ string at the final time step: 
\begin{equation}
\label{eq: cluster_weights} 
\begin{aligned}
\braket{\ec}{\psi^{\otimes 4}} &= \frac{1}{q}\\
\braket{\fc}{\psi^{\otimes 4}} &= \frac{1}{q(q+1)},
\end{aligned}
\end{equation}
where $\fc$ is discouraged by a factor of $\frac{1}{q+1}$ relative to $\ec$. We absorb the common $\frac{1}{q}$ weight with the $q^{L-1}$ factor. The result is 
\begin{equation}
\begin{aligned}
 &\overline{|\bra{\psi} O(t) \ket{\psi}|^2}  = \tr(O^2) /q \\
&\sum_{p \in \{\ec, \fc \}^{\otimes L}} \left(\frac{1}{q+1}\right)^{\# \text{of} \fc \text{in } p }  \langle p^* | \prod_{\tau = 1}^t\mathcal{T}(\tau) \ket{\fc \ec \cdots \ec}.
\end{aligned}
\end{equation}

The normalization convention is determined by the fact that when $O(t) = \I$, the top boundary state is $\ket{\ec \ec \cdots \ec}$ and the result should be $1$, since $\langle \psi |\psi \rangle = 1$.

The decay rate of $\overline{|\bra{\psi} O(t)\ket{\psi}|^2}$ can be understood from the emergent modes in the cluster picture. In a generic random circuit, the endpoint of a cluster undergoes a biased random walk with the butterfly velocity $v_B$ as its drift velocity, which is 1 for dual unitary circuits. Therefore, the boundary $\fc$ will travel uni-dirctionally with velocity $1$ on the light cone \cite{Claeys_2020, Foligno_Bertini_2023, Zhou_Harrow_2022}. If the bound state is favored, the two random walkers must travel on the same light cone. Only the exchange process $\ec \fc \leftrightarrow \ec \fc$ contributes to the propagation of the bound state on the light cone, and the transition probability will yield the magnon rate through $- \ln \mathbb{P}(\ec \fc \to \ec \fc) / \ln (2) = \ln \frac{3}{2-\cos(\pi a_z)} / \ln (2)$. 

The decay rate of the unbound state is less clear. Nonetheless, we expect a single-site operator to evolve to a random operator within the light cone, in the sense that it has a probability $\frac{1}{q^2}$ of being identity and or any one of the $q^2-1$ Paulis. The state $\ket{\fc \ec \cdots \ec}$ under the random dual unitary circuit evolution is expected to be statistically similar to the following random state after $t$ steps ($t \le L$):
\begin{equation}
\ket{\rm rand} = \left(\frac{1}{q^2} \ket{\ec} + \frac{q^2-1}{q^2} \ket{\fc} \right)^{\otimes t}\ket{\fc}  \ket{\ec}^{\otimes L-t}.
\end{equation}
As a result, the one-point function squared is 
\begin{equation}
    \begin{aligned}
    &\overline{|\bra{\psi} O(t) \ket{\psi}|^2}  \sim \braket{\psi^{\otimes 4}}{\rm rand} \\
    & = \frac{1}{q+1} \left( \frac{1 }{q^2} + \frac{q^2 - 1}{q^2} \frac{1}{q+1} \right)^t= \frac{1}{q+1}  e^{- t \ln q}. 
    \end{aligned}
\end{equation}
The exponent shows that the decay rate is $1$, equal to the domain wall rate in the spin basis. 

\begin{figure}
    \centering
    \includegraphics[width=1\linewidth]{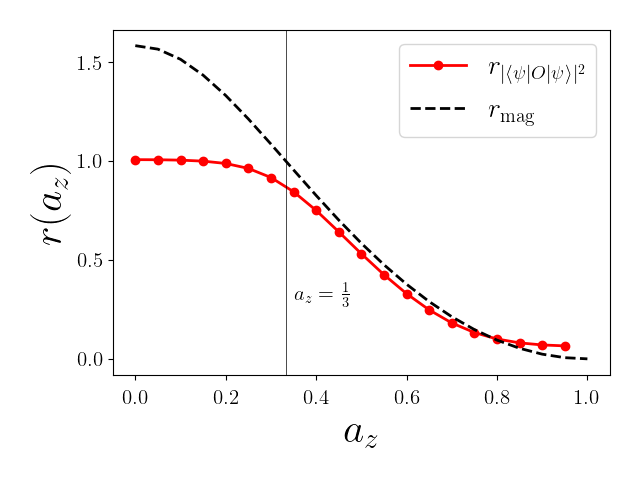}
    \caption{The second-stage decay rate of the one-point function squared computed in the cluster basis for the OBC in \bw geometry. The system size is $L=16$.}
    \label{fig: one-pt decay}
\end{figure}

\subsection{OTOC}

In this section, we carry out the cluster basis formalism to the OTOC. As hinted in the one-point function squared, the dominant state in one phase is a linear superposition in the cluster basis. So the dominant trajectory which used to be continuous snapshots of classical configurations also becomes a superposition, making it difficult to compute the decay rates. We conclude that the basis is not a good choice when domain walls appear in the trajectory. 

From the definition of $\{\ket{\ec}, \ket{\fc}\}$, the following identities allow us to directly translate between the spin basis and the cluster basis:
\begin{equation}
\begin{aligned}
|+ \rangle  = q | \ec \rangle,  &\quad | - \rangle  = | \ec \rangle  + (q^2 - 1) | \fc \rangle, \\
|-^* \rangle  = | \fc \rangle,  &\quad | +^* \rangle  = \frac{1}{q} ( | \ec \rangle  - | \fc \rangle  ).
\end{aligned}
\end{equation}
The OTC evolves from the same initial state
\begin{equation}
\label{eq:OTOC_in_cluster}
\begin{aligned}
    & \quad \overline{\langle V W(t)V W(t) \rangle} =  \cancelto{\text{set to }1}{\Tr(V^2) \Tr(W^2) \frac{1}{q} }\\
    &\frac{1}{q^{L-1}} \langle  - \cdots +^* \cdots - | \prod_{t'=1}^t \mathcal{M}(t') | -^* + \cdots + \rangle  \\
   & =  \langle  - \cdots +^* \cdots - | \prod_{t'=1}^t \mathcal{T}(t') \ket {\fc \ec \cdots  \ec}. 
\end{aligned}
\end{equation}
Were we to repeat the exercise of inferring the decay rate by searching the dominant trajectory, we need to generalize the concept of the dominant state to the probability distribution that has the largest overlap with the boundary state.

In the spin basis, the final state $\bra{- \cdots - +^* - \cdots -}$ fixes the spin to be $+$ at the site of $+*$ and favors $-$ spins by a factor of $q$ on each site elsewhere. The dominant state therefore is a $+$ magnon with $\mathcal{O}(1)$ width around $+^*$. The width, which represents superpositions of similar magnon with $\mathcal{O}(1)$ positional uncertainties about its $+$ core does not change the rate in the leading order. The boundary state constraints are different in the cluster basis. Based on \eqref{eq: cluster_normalization}. The site occupied by $+^*$ favors $\ec$, since $\langle  +^* | \ec \rangle  = \frac{1}{q}$, $\langle  +^* | \fc \rangle = - \frac{1}{q} \frac{1}{q^2 - 1}$. More importantly, the $-$ treats the two states on an equal footing due to equal overlap $\langle - | \ec \rangle  = \langle  - | \fc \rangle  = 1$. Therefore the dominant state can be a superposition of exponentially many states, with almost equal amplitudes. 

\begin{figure}
    \centering
    \includegraphics[width=1\linewidth]{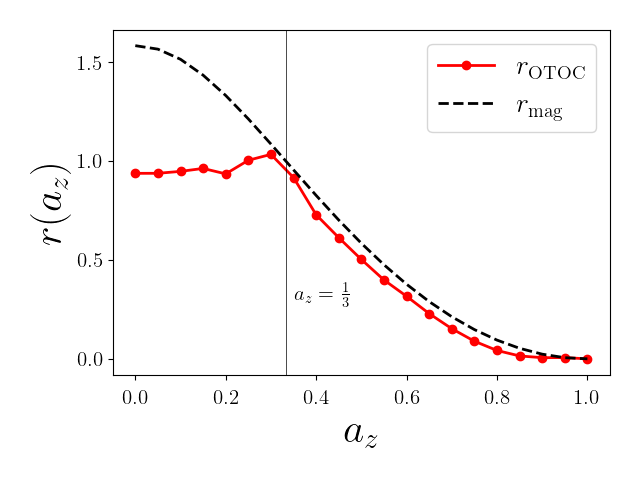}
    \caption{The second-stage decay rates of the OTOC computed in the cluster basis in \bw  geometry for OBC. $L=16$.}
    \label{fig: otoc_decay_cluster}
\end{figure}

\begin{figure}
    \centering
    \includegraphics[width=1\linewidth]{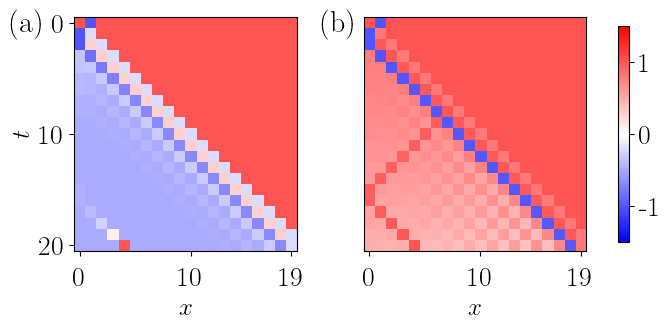}
    \caption{OTOC relaxation in the cluster basis for (a) $a_z = 0.2$ (b) $a_z = 0.7$, where $+1$/red represents unoccupied sites and $-1$/blue represents occupied sites.}
    \label{fig: otoc_cluster}
\end{figure}

Nonetheless, the OTOC can be calculated by setting the final boundary condition with the on-site probability $(p_{\ec},p_{\fc} )$ to be $(\frac{1}{2}, \frac{1}{2})$ at the $-$ spots and $(\frac{1}{q}, -\frac{1}{q(q+1)})$ at the $+^*$ spot. The OTOC also shows a two-stage decay in the cluster basis (Fig.~\ref{fig: otoc_decay_cluster}). Though the dominant trajectory cannot be found through maximization over classical intermediate configurations, the magnetization averaged over the trajectories with the boundary conditions specified by OTOC can be informative. Setting the polarization of $\ec$ to be 1 and $\fc$ to be $-1$, we see that its heatmap (Fig.~\ref{fig: otoc_cluster}) is similar to Fig.~\ref{fig: BW stage 1}, but the trajectory that leads to $r_{\rm mag}/2$ is much less obvious. As mentioned before, the states preferred by the bra include all the configurations with arbitrary combinations of $\ec$ and $\fc$ on all sites except $x_f = \ec$ at time $t$, so no single classical configuration stands out as the dominant one.

\section{Conclusion}

The study of OTOC and quantum chaos has often centered on the quantum butterfly effect, namely how a local perturbation spreads through scrambling in the operator space. Here, we focus on the relaxation that follows when this local signal has reached the probe. In a finite system, the OTOC’s exponential decay to its saturation value exhibits two stages, with the transition time proportional to $L / v_B$. Building on our previous work on the emergent modes explaining the two-stage decay of the half-system purity, we extend this theory to show that the same emergent modes govern the two-stage decay of the OTOC in an effective statistical model. 

In a solvable one-parameter family of dual-unitary circuits, we use a random averaging technique to introduce an effective statistical model with domain-wall and magnon modes. We find that the first-stage decay rate is half the magnon rate due to a geometric constraint in which the magnon traverses only half of the dominant trajectory. The second-stage decay rate arises from a competition between the domain wall and magnon mode, similar to the second stage observed in the purity decay during entanglement growth. While periodic boundary conditions and specific circuit geometries, such as a staircase configuration, introduce slight modifications, the essential structure of the decay remains intact. We validate these results numerically, reveal the dominant trajectories through heatmaps of ``magnetization" in the spin model, and confirm our analytic predictions using MPS numerics in large systems. 

Our findings also extend to chaotic Floquet circuits without disorder. In these models, we cannot restrict the problem to the two spin degrees of freedom in the Markovian dynamics since there is no random averaging. Instead, we directly simulate the unitary dynamics to compute the OTOC. We confirm that the magnon mode determines the decay rates in a similar manner as in the effective model %
when the system parameters allow for a clear two-stage decay. However, for certain parameter regimes in the clean Floquet model, the system may fully thermalize after the first stage, without exhibiting a second stage. In other regions of parameter space, we observe a strikingly slow decay in the second stage, with characteristics similar to a power-law behavior. We have not yet identified the precise conditions that give rise to this very slow second-stage decay.

We further analyze the decay using the cluster picture, where the OTOC is represented by a Pauli string basis that quantifies operator support. In this picture, emergent modes appear as bound states of clusters and random operators. They yield correct decay rates for the one-point function squared in the dual unitary circuit family. However, due to the superposition of configurations in the cluster basis, it is not straightforward to visualize the dominant trajectories. Nonetheless, numerical results remain consistent with our theoretical predictions. 

Finally, we list a few directions for future exploration. 

The cluster picture gives a binding transition for the two rates in the one-point function squared. However, one emergent mode -- a random operator -- is a superposition of classical configuration, which prevents us from further viewing the unbinding transition in a simple classical picture. It would be insightful to investigate further whether a local observable can yield a heatmap as effectively as the spin basis. We are also interested in the connection between the unbinding transition here and the one observed in the two-point function squared\cite{nahum_real-time_2022-1}. 

The unitary evolution we consider here has no conserved quantity. It would be valuable to understand how conserved charges, for instance, the energy in an analog simulator affect the two-stage decay. The slow diffusive charge persists for a much longer time scale than the butterfly wave. It may completely shadow the two-stage decay. We can also consider the effect of the dynamical constraints, such as those in the Rydberg atom arrays. 

At a small value of  $x$ , the first stage of decay occurs immediately with a decay rate $r_{\rm mag} / 2$, which is smaller than either the purity or one-point function squared decay rate. The OTOC thus provides a better and stronger signal to measure the magnon rate in an experiment. We can also explore higher-order OTOC, which involves more than 4 operators. The spin basis can contain more spins than $+/-$ and can host more modes. It would be interesting to see the competition of those modes in higher-order purity and OTOC. 

Noise is unavoidable in an experiment, and can affect the OTOC decay significantly. As the simplest example, a depolarization noise always prefers the $+$ spin over the $-$. It thus can be modeled as a global magnetic field that favors $+$. Then configurations with a $-$ magnon will be at a great disadvantage. It would be interesting to how the dominant trajectories change in the weak noise and strong noise regimes respectively. 

\textbf{Acknowledgements} CJ acknowledges funding from the European Union HORIZON-CL4-2022-QUANTUM-02-SGA through PASQuanS2.1 (Grant Agreement No. 101113690). TZ acknowledges the support from the physics department of Virginia Tech. 
\appendix

\bibliography{bib}
\bibliographystyle{unsrt}
\end{document}

%% file: tikz_lib.tex
\usepackage{xifthen}
\usepackage{xargs}

\newcommandx{\drawbox}[6][1=0,2=0,3=1,4=1,5=,6=]{
  \ifthenelse{\equal{#5}{}}{
    \draw[line width = 0.5pt] (#1,#2) rectangle (#3,#4);
  }{
    \draw[line width = 0.5pt, fill = #5] (#1,#2) rectangle (#3,#4);
  }
  \node () at (#1*0.5+#3*0.5,#2*0.5+#4*0.5) {#6};
}

\newcommandx{\regbox}[4][1=0,2=0,3=,4=]{
  \begin{scope}[shift={(#1,#2)}]
    \drawbox[0][0][1.5][1][#3][#4];
  \end{scope}
}

\newcommandx{\regboxt}[8][1=0,2=0,3=,4=,5=,6=,7=,8=]{
  \begin{scope}[shift={(#1,#2)}]
    \drawbox[0][0][1.5][1][#3][#4];
    \node[above] () at (0,1) {#5};
    \node[above] () at (1.5,1) {#6};
    \node[below] () at (1.5,0) {#7};
    \node[below] () at (0,0) {#8};
  \end{scope}
}

\newcommandx{\sqzbox}[5][1=0,2=0,3=,4=,5=]{
  \begin{scope}[shift={(#1,#2)}]
    \drawbox[0][0][2.25][1][#3][#4];
    \ifthenelse{\equal{#5}{s}}{
      \draw (0,0.1)--++(-0.1,0)--++(0,1)--++(2.25,0)--++(0,-0.1);
    }{}
  \end{scope}
}

\newcommandx{\ballon}[5][1=0,2=0,3=0.25,4=0.5,5=]{
  \begin{scope}[shift={(#1,#2)}]
    \pgfmathsetmacro{\r}{#4};
    \pgfmathsetmacro{\cent}{#3+#4};
    \draw (0,0)--++(0,#3);
    \draw (0,\cent) circle (\r);
    \draw (0,\cent+\r)--++(0,#3);
    \ifthenelse{\equal{#5}{}}{}{
      \node () at (0,\cent) {#5};
    }
  \end{scope}
}

\newcommandx{\wallebox}[5][1=,2=,3=,4=,5=]
{
  \fineq[-0.8ex][0.8][0.8]{
    \sqzbox[0][0];
    \ifthenelse{\equal{#1}{t}}{
      \ballon[0.5][1][0.25][0.5][$#2$];
      \ballon[1.75][1][0.25][0.5][$#3$];
    }{}
    \ifthenelse{\equal{#1}{b}}{
      \ballon[0.5][-1][0.25][0.5][$#4$];
      \ballon[1.75][-1][0.25][0.5][$#5$];
    }{}
    \ifthenelse{\equal{#1}{tb}}{
      \ballon[0.5][1][0.25][0.5][$#2$];
      \ballon[1.75][1][0.25][0.5][$#3$];
      \ballon[0.5][-1.5][0.25][0.5][$#4$];
      \ballon[1.75][-1.5][0.25][0.5][$#5$];
    }{}
  }
}

\newcommandx{\sufourpart}[5][1=0,2=0,3=0.25,4=0.5,5=]{
  \begin{scope}[shift={(#1,#2)}]
    \pgfmathsetmacro{\r}{#4};
    \pgfmathsetmacro{\cent}{#3+#4};
    \draw (0,0)--++(0,#3);
    \draw (-\r,\cent-\r) rectangle (\r,\cent+\r);
    \draw (0,\cent+\r)--++(0,#3);
    \ifthenelse{\equal{#5}{}}{}{
      \node () at (0,\cent) {#5};
    }
  \end{scope}
}

\newcommandx{\sufour}[5][1=,2=,3=,4=,5=]
{
  \fineq[-0.8ex][0.6][1]{
    \sqzbox[0][0][][$#1$];
    \sufourpart[0.5][1][0.25][0.5][$#2$];
    \sufourpart[1.75][1][0.25][0.5][$#3$];
    \sufourpart[0.5][-1.5][0.25][0.5][$#4$];
    \sufourpart[1.75][-1.5][0.25][0.5][$#5$];
  }
}

\newcommandx{\tisingu}[4][1=,2=,3=,4=]
{
  \fineq[-0.8ex][0.6][1]{
    \sqzbox[0][-1][][$#1$];
    \sqzbox[0][1.5][][$#1$];
    \sufourpart[0.5][0][0.25][0.5][$#2$];
    \sufourpart[1.75][0][0.25][0.5][$#3$];
    \draw (0.5,2.5)--++(0,0.25);
    \draw (1.75,2.5)--++(0,0.25);
    \draw (0.5,-1)--++(0,-0.25);
    \draw (1.75,-1)--++(0,-0.25);
  }
}

\newcommandx{\squbox}[4][1=0,2=0,3=,4=]{
  \begin{scope}[shift={(#1,#2)}]
    \drawbox[0][0][1][1][#3][#4];
  \end{scope}
}

\newcommandx{\uonesite}[7][1=0,2=0,3=,4=,5=,6=,7=]{
  \begin{scope}[shift={(#1,#2)}]
    \draw[line width = 0.5pt] (0.5,0.25)--++(0,-0.25) coordinate (A);
    \draw[line width = 0.5pt] (0.5,1.25)--++(0,0.25) coordinate (B);
    \ifthenelse{\equal{#7}{s}}{
      \squbox[0][0.25][#3][#4];
    }{
      \regbox[-0.25][0.25][#3][#4];      
    }
    \ifthenelse{\equal{#5}{}}{}{
      \node[below] () at (A) {#5};
    }
    \ifthenelse{\equal{#6}{}}{}{
      \node[above] () at (B) {#6};
    }
  \end{scope}
}

\newcommandx{\ugate}[5][1=0,2=0,3=,4=,5=]{
  \begin{scope}[shift={(#1,#2)}]
      \draw[line width = 0.5pt] (0,0)--++(0,0.25);
      \draw[line width = 0.5pt] (0,1.25)--++(0,0.25);
      \draw[line width = 0.5pt] (1,0)--++(0,0.25);
      \draw[line width = 0.5pt] (1,1.25)--++(0,0.25);
      \regbox[-0.25][0.25][#3][#4];
      \ifthenelse{\equal{#5}{p}}{
        \draw[line width = 2pt] (-0.1,0)--(1.1,0);
        \draw (0,0)--++(0,-0.1);
        \draw (1,0)--++(0,-0.1);
      }{}
  \end{scope}
}

\newcommandx{\dualgate}[7][1=0,2=0,3=,4=,5=,6=,7=]{
  \begin{scope}[shift={(#1,#2)}]
    \ifthenelse{\equal{#3}{l} \OR \equal{#3}{}}{
      \draw (0,-0.2)--(0.2,-0.2)--(0.2,0.2)--(0,0.2);
      \draw (0.2,0.2)--(0.5,0.5);
      \node () at (0.6,0.6) {$#5$};
      \draw (0.2,-0.2)--(0.5,-0.5);
      \node () at (0.6,-0.6) {$#7$};
    }{}
    \ifthenelse{\equal{#3}{r} \OR \equal{#3}{}}{
      \draw (0,-0.2)--(-0.2,-0.2)--(-0.2,0.2)--(0,0.2);
      \draw (-0.2,0.2)--(-0.5,0.5);
      \node () at (-0.6,0.6) {$#4$};
      \draw (-0.2,-0.2)--(-0.5,-0.5);
      \node () at (-0.6,-0.6) {$#6$};
    }{}
  \end{scope}
}

\newcommandx{\shortarc}[4][1=0,2=0,3=,4=]{
  \begin{scope}[shift={(#1,#2)}]
    \ifthenelse{\equal{#3}{r}}{
      \draw[line width = 0.5pt] (0,0) to[out=-90,in=0] (-0.1+0.06,0.1-0.18) to[out=180,in=-90] (-0.1,0.1);
      \pgfmathsetmacro{\flag}{-1};
    }{
      \draw[line width = 0.5pt] (0,0) to[out=90,in=0] (-0.06,0.18) to[out=180,in=90] (-0.1,0.1);
      \pgfmathsetmacro{\flag}{1};
    }
    \ifthenelse{\equal{#4}{l}}{
      \draw (0,0)--++(0,-0.2);
      \draw (-0.1,0.1)--++(0,-0.2);
    }{}
  \end{scope}
}

\newcommandx{\longarc}[4][1=0,2=0,3=,4=]{
  \begin{scope}[shift={(#1,#2)}]
    \ifthenelse{\equal{#3}{r}}{
      \draw[line width = 0.5pt] (0,0) to[out=-90,in=0] (-0.1*3+0.06*3,0.1*3-0.14*3) to[out=180,in=-90] (-0.1*3,0.1*3);
      \pgfmathsetmacro{\flag}{-1};
    }{
      \draw[line width = 0.5pt] (0,0) to[out=90,in=0] (-0.06*3,0.14*3) to[out=180,in=90] (-0.1*3,0.1*3);
      \pgfmathsetmacro{\flag}{1};
    }
    \ifthenelse{\equal{#4}{l}}{
      \draw (0,0)--++(0,-0.2);
      \draw (-0.1*3,0.1*3)--++(0,-0.2);
    }{}
  \end{scope}
}

\newcommandx{\idarc}[4][1=0,2=0,3=,4=]{
  \begin{scope}[shift={(#1,#2)}]
    \shortarc[0][0][#3][#4];
    \shortarc[-0.2][0.2][#3][#4];
  \end{scope}
}

\newcommandx{\swaparc}[4][1=0,2=0,3=,4=]{
  \begin{scope}[shift={(#1,#2)}]
    \longarc[0][0][#3][#4];
    \shortarc[-0.1][0.1][#3][#4];
  \end{scope}
}

\newcommandx{\idst}[3][1=0,2=0,3=]{
  \ifthenelse{\equal{#3}{r}}{
    \pgfmathsetmacro{\flag}{-1};
  }{
    \pgfmathsetmacro{\flag}{1};
  }
  \begin{scope}[shift={(#1,#2)}]
    \draw[line width = 0.5pt] (0,0) to[out=\flag*90,in=180] (0.15,\flag*0.2) to[out=0,in=\flag*90] (0.3,0);
    \draw[line width = 0.5pt] (0.4,0) to[out=\flag*90,in=180] (0.4+0.15,\flag*0.2) to[out=0,in=\flag*90] (0.7,0);
  \end{scope}
}

\newcommandx{\swapst}[3][1=0,2=0,3=]{
  \ifthenelse{\equal{#3}{r}}{
    \pgfmathsetmacro{\flag}{-1};
  }{
    \pgfmathsetmacro{\flag}{1};
  }
  \begin{scope}[shift={(#1,#2)}]
      \draw[line width = 0.5pt] (0,0) to[out=\flag*90,in=180] (0.4,\flag*0.25) to[out=0,in=\flag*90] (0.8,0);
      \draw[line width = 0.5pt] (0.3,0) to[out=\flag*90,in=180] (0.4,\flag*0.15) to[out=0,in=\flag*90] (0.5,0);
  \end{scope}
}

\newcommandx{\uonesitestack}[4][1=,2=,3=1,4=]
{
  \begin{scope}[shift={(#1,#2)}]
    \foreach \x in {#3,...,1}{
      \pgfmathsetmacro{\shiftx}{-0.1*(2*\x-2)};
      \pgfmathsetmacro{\shifty}{0.1*(2*\x-2)};
      \uonesite[\shiftx-0.1][\shifty+0.1][blue!50];
      \uonesite[\shiftx][\shifty][red!50];
    }
    \ifthenelse{\equal{#4}{id}}{
      \idarc[0.5][1.5];
    }{}
    \ifthenelse{\equal{#4}{swap}}{
      \swaparc[0.5][1.5];
    }{}
  \end{scope}
}
\newcommandx{\ugatestack}[6][1=0,2=0,3=1,4=,5=,6=]
{
  \begin{scope}[shift={(#1,#2)}]
    \foreach \x in {#3,...,1}{
      \pgfmathsetmacro{\shiftx}{-0.1*(2*\x-2)};
      \pgfmathsetmacro{\shifty}{0.1*(2*\x-2)};
      \ifthenelse{\equal{#6}{l}}{
        \ugatel[\shiftx-0.1][\shifty+0.1][][blue!50];
        \ugatel[\shiftx][\shifty][][red!50];
      }{}
      \ifthenelse{\equal{#6}{r}}{
        \ugater[\shiftx-0.1][\shifty+0.1][][blue!50];
        \ugater[\shiftx][\shifty][][red!50];
      }{}
      \ifthenelse{\equal{#6}{}}{
        \ugate[\shiftx-0.1][\shifty+0.1][blue!50];
        \ugate[\shiftx][\shifty][red!50];
      }{}
    }
    \ifthenelse{\equal{#4}{id}}{
      \idarc[0][1.5];
    }{}
    \ifthenelse{\equal{#4}{swap}}{
      \swaparc[0][1.5];
    }{}
    \ifthenelse{\equal{#5}{id}}{
      \idarc[1][1.5];
    }{}
    \ifthenelse{\equal{#5}{swap}}{
      \swaparc[1][1.5];
    }{}
  \end{scope}
}

\newcommandx{\dptri}[7][1=0,2=0,3=,4=,5=,6=,7=]{
  \begin{scope}[shift={(#1,#2)}]
    \ifthenelse{\equal{#3}{}}{
      \draw (0,0)--(-1,1.732)--(1,1.732)--cycle;
    }{
      \draw[fill=#3] (0,0)--(-1,1.732)--(1,1.732)--cycle;
    }
    \node () at (0,1.15) {#4};
    \node[below] () at (0,0) {#5};
    \node[left] () at (-1,1.732) {#6};
    \node[right] () at (1,1.732) {#7};
  \end{scope}
}

\newcommandx{\dptridash}[7][1=0,2=0,3=,4=,5=,6=,7=]{
  \begin{scope}[shift={(#1,#2)}]
    \dptri[0][0][#3][#4][#5][#6][#7]
    \draw[dashed] (0,0)--++(0,1.732*2/3)--(0-1,0+1.732);
    \draw[dashed] (0,1.732*2/3)--(0+1,0+1.732);
  \end{scope}
}

\newcommandx{\dtptri}[4][1=0,2=0,3=,4=]
{
  \begin{scope}[shift={(#1,#2)}]
    \draw (0,0)--(-1,1.732)--(1,1.732)--cycle;
    \draw (0,0)--(-1,-1.732)--(1,-1.732)--cycle;
  \end{scope}
}

\newcommandx{\fineq}[4][1=-.8ex,2=1,3=1]{
  \begin{tikzpicture}[baseline={([yshift=#1]current  bounding  box.center)}, scale = #2, every node/.style={scale = #3}]
    #4
  \end{tikzpicture}
}

\newcommandx{\ideq}[1][1=]{
  \fineq{
    \idst[0][0][#1]
  }
}
\newcommandx{\swapeq}[1][1=]{
  \fineq{
    \swapst[0][0][#1]
  }
}

\newcommandx{\idket}[0]{
  |
  \fineq[-0.6ex]{
    \idst[0][0][r]
  } \rangle 
}

\newcommandx{\swapket}[0]{
  |
  \fineq[-0.4ex]{
    \swapst[0][0][r]
  } \rangle 
}

\newcommandx{\idbra}[0]{
  \langle 
  \fineq{
    \idst[0][0][]
  } |
}

\newcommandx{\swapbra}[0]{
  \langle 
  \fineq{
    \swapst[0][0][]
  } |
}

\newcommandx{\smallarc}[2][1=,2=]{
  \ifthenelse{\equal{#2}{r}}{
    \pgfmathsetmacro{\flag}{-1};
  }{
    \pgfmathsetmacro{\flag}{1};
  }
  \ifthenelse{\equal{#1}{}}{
    \draw[line width = 0.5pt] (0,0) to[out=\flag*90,in=180] (0.5,\flag*0.5) to[out=0,in=\flag*90] (1,0);
  }{
    \draw[line width = 0.5pt] (0,0) to[out=\flag*90,in=190] (0.5-0.2,\flag*0.45);
    \draw[line width = 0.5pt] (0.5+0.2,\flag*0.45) to[out=-10,in=\flag*90] (1,0);
    \node () at (0.5,\flag*0.45) {#1};
  }
}

\newcommandx{\oidst}[5][1=0,2=0,3=,4=,5=]{
  \begin{scope}[shift={(#1,#2)}]
    \smallarc[#3][#5]
    \begin{scope}[shift={(1.5,0)}]
      \smallarc[#4][#5]
    \end{scope}
  \end{scope}
}

\newcommandx{\higharc}[2][1=,2=]{
  \ifthenelse{\equal{#2}{r}}{
    \pgfmathsetmacro{\flag}{-1};
  }{
    \pgfmathsetmacro{\flag}{1};
  }
  \ifthenelse{\equal{#1}{}}{
    \draw[line width = 0.5pt] (0,0) to[out=\flag*90,in=180] (1,\flag*0.5) to[out=0,in=\flag*90] (2,0);
  }{
    \draw[line width = 0.5pt] (0,0) to[out=\flag*90,in=180] (1-0.2,\flag*0.55);
    \draw[line width = 0.5pt] (1+0.2,\flag*0.55) to[out=0,in=\flag*90] (2,0);
    \node () at (1,\flag*0.55) {#1};
  }
}

\newcommandx{\lowarc}[2][1=,2=]{
  \ifthenelse{\equal{#2}{r}}{
    \pgfmathsetmacro{\flag}{-1};
  }{
    \pgfmathsetmacro{\flag}{1};
  }
  \ifthenelse{\equal{#1}{}}{
    \draw[line width = 0.5pt] (0,0) to[out=\flag*90,in=180] (0.3,\flag*0.3) to[out=0,in=\flag*90] (0.6,0);
  }{
    \draw[line width = 0.5pt] (0,0) to[out=\flag*90,in=190] (0.3-0.2,\flag*0.28);
    \draw[line width = 0.5pt] (0.3+0.2,\flag*0.28) to[out=-10,in=\flag*90] (0.6,0);
    \node () at (0.3,\flag*0.28) {#1};
  }
}

\newcommandx{\oswapst}[5][1=0,2=0,3=,4=,5=]{
  \begin{scope}[shift={(#1,#2)}]
    \higharc[#3][#5]
    \begin{scope}[shift={(0.75,0)}]
      \lowarc[#4][#5]
    \end{scope}
  \end{scope}
}

\newcommandx{\uonesiteeq}[3][1=,2=,3=]
{
  \fineq{
    \uonesite[0][0][][#1][#2][#3];
    \node () at (-0.5,0.75) {};
    \node () at (1.5,0.75) {};
  }
}

\newcommandx{\ugateeq}[3][1=,2=,3=]
{
  \fineq{
    \ugate[0][0][][#1][#2][#3];
    \node () at (-0.5,0.75) {};
    \node () at (1.5,0.75) {};
  }
}

\newcommandx{\uonesitetwoeq}[4][1=,2=,3=,4=]
{
  \fineq{
    \uonesite[0][0][][#1][#3];
    \uonesite[0][1.5][][#2][#4];
    \node () at (-0.5,0.75) {};
    \node () at (1.5,0.75) {};
  }
}

\newcommandx{\hfbox}[5][1=0,2=0,3=,4=l,5=]{
  \ifthenelse{\equal{#4}{r}}{
    \pgfmathsetmacro{\flag}{-1};
  }{
    \pgfmathsetmacro{\flag}{1};
  }
  \begin{scope}[shift={(#1,#2)}]
    \draw (0,0) --++ (\flag*1,0) --++ (0,1) --++(-\flag*1,0);
    \node () at (\flag*0.5,0.5) {#3};
    \ifthenelse{\equal{#5}{s}}{
      \ifthenelse{\equal{#4}{l}}{
        \draw (0.9,1)--++(0,0.1) --++ (-1,0);
      }{}
      \ifthenelse{\equal{#4}{r}}{
        \draw (-1,0.1)--++(-0.1,0)--++(0,1) --++ (1,0);
      }{}
    }{}
  \end{scope}
}

\newcommandx{\tribox}[4][1=,2=,3=,4=]{
  \fineq[-0.8ex][0.35][0.8]{
    \hfbox[0][1.25][$#1$][l][#4];
    \hfbox[2.25][1.25][$#2$][r][#4];
    \sqzbox[0][0][][$#3$][#4];
    \node () at (-0.25,1.25) {};
    \node () at (2.5,1.25) {};
  }
}

\newcommandx{\lshapebox}[4][1=,2=,3=l,4=]{
  \fineq[-0.8ex][0.35][0.55]{
    \ifthenelse{\equal{#3}{l}}{
      \hfbox[0][1.25][$#1$][l][#4];
    }{
      \hfbox[2.25][1.25][$#1$][r][#4];
    }
    \sqzbox[0][0][][$#2$][#4];
    \node () at (-0.25,1.25) {};
    \node () at (2.5,1.25) {};
  }
}

\newcommandx{\hexbox}[8][1=,2=,3=,4=,5=,6=,7=,8=]{
  \fineq[-0.8ex][0.3][0.48]{
    \sqzbox[0][2.5][][$#1$][#8];
    \sqzbox[2.5][2.5][][$#2$][#8];
    \hfbox[0][1.25][$#3$][l][#8];
    \sqzbox[1.25][1.25][][$#4$][#8];
    \hfbox[2.5+2.25][1.25][$#5$][r][#8];
    \sqzbox[0][0][][$#6$][#8];
    \sqzbox[2.5][0][][$#7$][#8];
    \node () at (-0.25,1.25) {};
    \node () at (5,1.25) {};
  }
}

\newcommandx{\regboxteq}[5][1=,2=,3=,4=,5=]{
  \fineq[-0.8ex][0.5][0.7]{
    \regboxt[0][0][][#1][$#2$][$#3$][$#4$][$#5$];
  }
}

\newcommandx{\enttworandqubit}[2][1=,2=]{
  \fineq[-0.8ex][0.6][0.5]{
    \sqzbox[0][0][][4U];
    \draw (0.5,1)--++(0,0.5) node[above] () {$+$};
    \draw (2.25-0.5,1)--++(0,0.5) node[above] () {$-$};
    \draw (0.5,0)--++(0,-0.5);
    \draw (2.25-0.5,0)--++(0,-0.5);
    \draw (0.5,-1.5)--++(0,-0.5) node[below] () {$e$};
    \draw (2.25-0.5,-1.5)--++(0,-0.5) node[below] () {$e$};
    \draw (0,-1.5) rectangle (1,-0.5);
    \draw (2.25-1,-1.5) rectangle (2.25,-0.5);
    \node () at (0.5,-1) {#1};
    \node () at (2.25-0.5,-1) {#2};
  }
}

\newcommandx{\ugatetext}[7][1=0,2=0,3=,4=,5=,6=,7=]
{
  \begin{scope}[shift={(#1,#2)}]
    \ugate[0][0][][#3];
    \node[above] () at (0,1.5) {$#4$};
    \node[above] () at (1,1.5) {$#5$};
    \node[below] () at (0,0) {$#6$};
    \node[below] () at (1,0) {$#7$};
  \end{scope}
}

\newcommandx{\stairgateone}[1][1=]{
  \fineq[-0.8ex][0.75][0.75]{
    \ugatetext[0][0][#1][+][-][\frac{+}{q^2}][\frac{-}{q^2}];
  }
}

\newcommandx{\stairgatebottom}[1][1=]{
  \fineq[-0.8ex][0.75][0.75]{
    \ugatetext[0][0][#1][+][][\frac{+}{q^2}][\frac{-}{q^2}];
  }
}

\newcommandx{\stairgatetwo}[2][1=,2=]{
  \fineq[-0.8ex][0.75][0.75]{
    \ugatetext[0][0][#1][+][][\frac{+}{q^2}][\frac{-}{q^2}];
    \ugatetext[1][1.5][#2][+][-][][\frac{-}{q^2}];
    \node () at (-0.5,1) {};
    \node () at (2.5,1) {};
  }
}

\newcommandx{\sqzboxwithantennae}[7][1=0,2=0,3=,4=,5=,6=,7=]
{
  \begin{scope}[shift={(#1,#2)}]
    \sqzbox[0][0][][#3];
    \ifthenelse{\equal{#4}{}}{
    }{
      \draw (0.625,1)--++(0,0.5) node[above] {#4};
    }
    \ifthenelse{\equal{#5}{}}{
    }{
      \draw (2.25-0.625,1)--++(0,0.5) node[above] {#5};
    }
    \ifthenelse{\equal{#6}{}}{
    }{
      \draw (0.625,0)--++(0,-0.5) node[below] {#6};
    }
    \ifthenelse{\equal{#7}{}}{
    }{
      \draw (2.25-0.625,0)--++(0,-0.5) node[below] {#7};
    }
  \end{scope}
  
}

\newcommandx{\pyramidone}[1][1=]{
  \fineq[-0.8ex][0.5][0.8]{
    \sqzboxwithantennae[0][0][#1][$+$][$-$];
    \node () at (-0.25,1.25) {};
    \node () at (2.5,1.25) {};
  }
}

\newcommandx{\pyramidtwo}[3][1=,2=,3=]{
  \fineq[-0.8ex][0.5][0.8]{
    \sqzboxwithantennae[1.25][1.25][#1][$+$][$-$];
    \sqzboxwithantennae[0][0][#2][$+$][];
    \sqzboxwithantennae[2.5][0][#3][][$-$];
    \node () at (-0.25,1.25) {};
    \node () at (5,1.25) {};
  }
}

\newcommandx{\pyramidthree}[6][1=,2=,3=,4=,5=,6=]{
  \fineq[-0.8ex][0.5][0.8]{
    \sqzboxwithantennae[2.5][2.5][#1][$+$][$-$];
    \sqzboxwithantennae[1.25][1.25][#2][$+$][];
    \sqzboxwithantennae[3.75][1.25][#3][][$-$];
    \sqzboxwithantennae[0][0][#4][$+$][];
    \sqzboxwithantennae[2.5][0][#5];
    \sqzboxwithantennae[5][0][#6][][$-$];
    \node () at (-0.25,1.25) {};
    \node () at (7.5,1.25) {};
  }
}

\newcommandx{\pyramidZone}[2][1=,2=]{
  \fineq[-0.8ex][0.4][0.64]{
    \sqzboxwithantennae[0][0][][$+$][$-$][$#1$][$#2$];
    \node () at (-0.25,1.25) {};
    \node () at (2.5,1.25) {};
  }
}

\newcommandx{\pyramidZtwo}[4][1=,2=,3=,4=,]{
  \fineq[-0.8ex][0.4][0.64]{
    \ifthenelse{\equal{#1}{l}}{
      \sqzboxwithantennae[1.25][1.25][][$+$][$-$][][$#4$];
      \sqzboxwithantennae[0][0][][$+$][][$#2$][$#3$];
    }{
      \ifthenelse{\equal{#1}{r}}{
        \sqzboxwithantennae[1.25][1.25][][$+$][$-$][$#2$];
        \sqzboxwithantennae[2.5][0][][][$-$][$#3$][$#4$];
      }{
        \sqzboxwithantennae[1.25][1.25][][$+$][$-$];
        \sqzboxwithantennae[0][0][][$+$][][$#1$][$#2$];
        \sqzboxwithantennae[2.5][0][][][$-$][$#3$][$#4$];
      }
    }
    \node () at (-0.25,1.25) {};
    \node () at (5,1.25) {};
  }
}

\newcommandx{\pyramidZttwo}[3][1=,2=,3=]{
  \fineq[-0.8ex][0.4][0.64]{
    \sqzboxwithantennae[2.5][2.5][][$+$][$-$];
    \sqzboxwithantennae[1.25][1.25][][$+$][];
    \sqzboxwithantennae[3.75][1.25][][][$-$];
    \hfbox[0][1.25][$+$][l];
    \sqzbox[0][0][][$#1$];
    \sqzboxwithantennae[2.5][0][$#2$];
    \hfbox[2.25+5][1.25][$-$][r];    
    \sqzbox[5][0][][$#3$];
    \node () at (-0.25,1.25) {};
    \node () at (7.5,1.25) {};
  }
}

\newcommandx{\stairgateany}[2][1=,2=]{
  \fineq[-0.8ex][0.75][0.75]{
    \ugatetext[0][0][#1][+][][\frac{+}{q^2}][\frac{-}{q^2}];
    \node () at (1.5,2.4) {$\iddots$};
    \draw (0.75,1.5)--++(0,0.5);
    \draw (0.75,1.5)--++(0.5,0);
    \draw (2.25,3)--++(0,-0.5);
    \draw (2.25,3)--++(-0.5,0);
    \ugatetext[2][3][#2][+][-][][\frac{-}{q^2}];
    \node () at (-0.5,1) {};
    \node () at (3.5,1) {};
  }
}

\newcommandx{\stairgatethree}[3][1=,2=,3=]{
  \fineq[-0.8ex][0.75][0.75]{
    \ugatetext[0][0][#1][+][][\frac{+}{q^2}][\frac{-}{q^2}];
    \ugatetext[1][1.5][#2][+][][][\frac{-}{q^2}];
    \ugatetext[2][3][#3][+][-][][\frac{-}{q^2}];
    \node () at (-0.5,1) {};
    \node () at (3.5,1) {};
  }
}

\newcommandx{\pingbox}[3][1=,2=,3=]{
  \fineq[-0.8ex][0.5][0.5]{
    \ugate[1][1.5][][#1];
    \ugate[0][0][][#2];
    \ugate[2][0][][#3];
    \foreach \x in {0,...,3}{
      \node[below] () at (\x,0) {$e$};
    }
    \draw[line width = 0.5pt] (0,1.5)--++(0,1.5);
    \draw[line width = 0.5pt] (3,1.5)--++(0,1.5);
    \node[above] () at (0,3) {$+$};
    \node[above] () at (1,3) {$+$};
    \node[above] () at (2,3) {$-$};
    \node[above] () at (3,3) {$-$};
  }
}

\newcommandx{\opsqu}[1][1=1]{
  \fineq[-0.8ex][0.2]{
    \ifthenelse{\equal{#1}{1}}{
      \draw[line width = 1pt] (0.125,0)--++(0,1);
      \draw[line width = 1pt] (0.875,0)--++(0,1);
    }{}
    \ifthenelse{\equal{#1}{2}}{
      \draw[line width = 1pt] (0,0)--++(1,1);
      \draw[line width = 1pt] (1,0)--++(-1,1);
    }{}
  }
}

\newcommandx{\bottomtri}[6][1=,2=,3=,4=,5=,6=]{
  \fineq[-0.8ex][0.4][0.6]{
    \dptri[0][0][][$#6$][][$#2$][$#3$];
    \node[right] () at (0,0) {$#1$};
    \draw (0,0)--++(-120:1) node[below left] () {$#4$};
    \draw (0,0)--++(-60:1) node[below right] () {$#5$};
  }
}

\newcommandx{\jabc}[4][1=,2=,3=,4=]
{
  \fineq[-0.8ex][0.4][0.8]{
    \dptri[0][0][][$#4$][$#1$][$#2$][$#3$];
  }
}

\newcommandx{\joneperp}[8][1=,2=,3=,4=,5=,6=,7=,8=]
{
  \fineq[-0.8ex][0.4][0.8]{
    \dptri[0][0][][$#4$][$#1$][$#2$][$#3$];
    \dptri[-1][-1.732][][$#6$][][$#5$][];
    \dptri[1][-1.732][][$#8$][][][$#7$];
  }
}

\newcommandx{\sdw}[1][1=l]
{
  \fineq[-0.4ex][0.5][1]{
    \dptri;
    \ifthenelse{\equal{#1}{l}}{
      \draw (0,1.155) -- +(90:1.077);
      \draw (0,1.155) -- +(210:1.077);
    }{}
    \ifthenelse{\equal{#1}{r}}{
      \draw (0,1.155) -- +(90:1.077);
      \draw (0,1.155) -- +(-30:1.077);
    }{}
    \ifthenelse{\equal{#1}{lp}}{
      \draw (0,1.2) -- +(90:1.077);
      \draw (0,1.2) -- +(210:1.077);
      \draw (0,1.1) -- +(-30:1.02);
      \draw (0,1.1) -- +(210:1.02);
    }{}
    \ifthenelse{\equal{#1}{lr}}{
      \draw (0,0)--(-1,-1.732)--(1,-1.732)--cycle;
      \draw (-0.05,1.732+0.5)--++(0,-0.5-1/3*1.732)--++(-1,-1/3*1.732)--++(0,-2/3*1.732)--++(1,-1/3*1.732)--++(0,-0.5-1/3*1.732);
    }{}
    \ifthenelse{\equal{#1}{rl}}{
      \draw (0,0)--(-1,-1.732)--(1,-1.732)--cycle;
      \draw (0.05,1.732+0.5)--++(0,-0.5-1/3*1.732)--++(1,-1/3*1.732)--++(0,-2/3*1.732)--++(-1,-1/3*1.732)--++(0,-0.5-1/3*1.732);
    }{}
    \node () at (-1.1,0) {};
    \node () at (1.1,0) {};
  }
}

\newcommandx{\sdwi}[1][1=]
{
  \fineq[-0.8ex][0.5][1]{
    \dptridash;
    \ifthenelse{\equal{#1}{l}}{
      \draw (0,1.732+0.5)--(0,1.732)--(-0.5,0.5*1.732)--++(-0.25*1.732,-0.25);
    }{}
    \ifthenelse{\equal{#1}{t}}{
      \draw (0,1.732+0.5)--(0,1.732)--(+0.5,0.5*1.732);
      \draw[red] (0.5, 0.5*1.732)--(-0.5,0.5*1.732);
      \draw (-0.5, 0.5*1.732)--++(-0.25*1.732,-0.25);
    }{}
    \ifthenelse{\equal{#1}{tri}}{
      \draw (0,1.732+0.5)--(0,1.732)--(-0.5,0.5*1.732)--++(-0.25*1.732,-0.25);
      \draw[red] (-0.4,0.5*1.732)--++(0.85,0)--++(120:0.85)--cycle;
    }{}
    \ifthenelse{\equal{#1}{lpd}}{
      \draw (0,1.732+0.5)--(0,1.732)--(-0.5,0.5*1.732)--++(-0.25*1.732,-0.25);
      \draw (-0.45,0.45*1.732)--++(60:0.9)--++(-60:0.9);
      \draw (-0.45,0.45*1.732)--++(-0.25*1.732,-0.25);
      \draw (0.45,0.45*1.732)--++(0.25*1.732,-0.25);
    }{}
    \ifthenelse{\equal{#1}{lpw}}{
      \draw (0,1.732+0.5)--(0,1.732)--(-0.5,0.5*1.732)--++(-0.25*1.732,-0.25);
      \draw (-0.45,0.45*1.732)--++(-0.25*1.732,-0.25);
      \draw (0.45,0.45*1.732)--++(0.25*1.732,-0.25);
      \draw[red] (-0.45,0.45*1.732)--(0.45,0.45*1.732);
    }{}
    \ifthenelse{\equal{#1}{lpt}}{
      \draw (0,1.732+0.5)--(0,1.732)--(+0.5,0.5*1.732);
      \draw[red] (0.5, 0.5*1.732)--(-0.5,0.5*1.732);
      \draw (-0.5, 0.5*1.732)--++(-0.25*1.732,-0.25);
      \draw (-0.45,0.45*1.732)--++(60:0.9)--++(-60:0.9);
      \draw (-0.45,0.45*1.732)--++(-0.25*1.732,-0.25);
      \draw (0.45,0.45*1.732)--++(0.25*1.732,-0.25);
    }{}
    \ifthenelse{\equal{#1}{lpwt}}{
      \draw (0,1.732+0.5)--(0,1.732)--(+0.5,0.5*1.732);
      \draw[red] (0.5, 0.5*1.732)--(-0.5,0.5*1.732);
      \draw (-0.5, 0.5*1.732)--++(-0.25*1.732,-0.25);
      \draw (-0.45,0.45*1.732)--++(-0.25*1.732,-0.25);
      \draw (0.45,0.45*1.732)--++(0.25*1.732,-0.25);
      \draw[red] (-0.45,0.45*1.732)--(0.45,0.45*1.732);
    }{}
    \node () at (-1.1,0) {};
    \node () at (1.1,0) {};
  }
}

\newcommandx{\ddw}[1][1=]
{
  \fineq[-0.4ex][0.5][1]{
    \dptri;
    \ifthenelse{\equal{#1}{l}}{
      \draw (0,1.155) -- +(90:1.077);
      \draw (0,1.155) -- +(210:1.077);
      \draw (-0.1,1.23) -- +(90:1.008);
      \draw (-0.1,1.23) -- +(210:1.02);  
    }{}
    \ifthenelse{\equal{#1}{lr}}{
      \draw (0.06,1.155) -- +(90:1.077);
      \draw (0.06,1.155) -- +(-30:1.077);
      \draw (-0.06,1.155) -- +(90:1.077);
      \draw (-0.06,1.155) -- +(210:1.077);
    }{}
    \ifthenelse{\equal{#1}{r}}{
      \draw (0,1.155) -- +(90:1.077);
      \draw (0,1.155) -- +(-30:1.077);
      \draw (0.1,1.23) -- +(90:1.008);
      \draw (0.1,1.23) -- +(-30:1.02);  
    }{}
    \ifthenelse{\equal{#1}{lp}}{
      \draw (0,1.04) -- +(-30:1.02);
      \draw (0,1.04) -- +(210:1.02);
      \draw (-0.1,1.732+0.5)--++(0,-0.45-1/3*1.732)--++(210:1.0);
      \draw (-0.0,1.732+0.5)--++(0,-0.5-1/3*1.732)--++(210:1.077);
    }{}
    \ifthenelse{\equal{#1}{lrp}}{
      \draw (0.05,1.732+0.5)--++(0,-0.5-1/3*1.732)--++(1,-1/3*1.732);
      \draw (-0.05,1.732+0.5)--++(0,-0.5-1/3*1.732)--++(-1,-1/3*1.732);
      \draw (0,0.616*1.732)--++(210:2/3*1.76);
      \draw (0,0.616*1.732)--++(-30:2/3*1.76);
    }{}
    \ifthenelse{\equal{#1}{lrpp}}{
      \draw (0.05,1.732+0.5)--++(0,-0.5-1/3*1.732)--++(1,-1/3*1.732);
      \draw (-0.05,1.732+0.5)--++(0,-0.5-1/3*1.732)--++(-1,-1/3*1.732);
      \draw (0,0.616*1.732)--++(210:2/3*1.76);
      \draw (0,0.616*1.732)--++(-30:2/3*1.76);
      \draw (0,0.54*1.732)--++(210:2/3*1.66);
      \draw (0,0.54*1.732)--++(-30:2/3*1.66);
    }{}
    \ifthenelse{\equal{#1}{regvert}}{
      \draw (0.05,1.732+0.5)--++(0,-0.5-1/3*1.732)--++(1,-1/3*1.732);
      \draw (-0.05,1.732+0.5)--++(0,-0.5-1/3*1.732)--++(-1,-1/3*1.732);
      \draw (0,0.616*1.732)--++(210:2/3*1.76) coordinate (A);
      \node[left] () at (A) {\footnotesize $(12)$};
      \draw (0,0.616*1.732)--++(-30:2/3*1.76) coordinate (B);
      \node[right] () at (B) {\footnotesize$(34)$};
      \node[above] () at (0,1.732+0.5) {\footnotesize $(12)(34)$};
      \node () at (-1-0.3,-1/1.732) {\footnotesize $\alpha$};
      \node () at (1+0.65,-1/1.732+0.1) {\footnotesize $\alpha^{-1}$};
      \draw[dashed] (A)--++(0,-2/3*1.55)--(0,-0.616*1.732);
      \draw[dashed] (B)--++(0,-2/3*1.55)--(0,-0.616*1.732);
    }{}
    \ifthenelse{\equal{#1}{specvert}}{
      \draw (0.05,1.732+0.5)--++(0,-0.5-1/3*1.732)--++(1,-1/3*1.732);
      \draw (-0.05,1.732+0.5)--++(0,-0.5-1/3*1.732)--++(-1,-1/3*1.732);
      \draw (0,0.616*1.732)--++(210:2/3*1.65);
      \draw (0,0.616*1.732)--++(-30:2/3*1.65);
      \fill (0,2/3*1.732-0.03) circle (0.1);
    }{}
    \node () at (-1.1,0) {};
    \node () at (1.1,0) {};
  }
}

\newcommandx{\ddwi}[1][1=]
{
  \fineq[-0.4ex][0.5][1]{
    \dptridash;
    \ifthenelse{\equal{#1}{l}}{
      \draw (0,1.732+0.5)--(0,1.732)--(-0.5,0.5*1.732)--++(-0.25*1.732,-0.25);
      \draw (-0.1,1.732+0.5)--(-0.1,1.732)--(-0.55,0.55*1.732)--++(-0.25*1.732,-0.25);
    }{}
    \ifthenelse{\equal{#1}{lr}}{
      \draw (-0.05,1.732+0.5)--++(0,-0.5)--(-0.55,0.55*1.732)--++(-0.25*1.732,-0.25);
      \draw (0.05,1.732+0.5)--++(0,-0.5)--(0.55,0.55*1.732)--++(0.25*1.732,-0.25);
    }{}
    \ifthenelse{\equal{#1}{r}}{
      \draw (0,1.732+0.5)--(0,1.732)--(0.5,0.5*1.732)--++(0.25*1.732,-0.25);
      \draw (0.1,1.732+0.5)--(0.1,1.732)--(0.55,0.55*1.732)--++(0.25*1.732,-0.25);
    }{}
    \ifthenelse{\equal{#1}{lrt}}{
      \draw (-0.05,1.732+0.5)--++(0,-0.5)--(-0.55,0.55*1.732)--++(-0.25*1.732,-0.25);
      \draw (0.05,1.732+0.5)--++(0,-0.5)--(-0.5,0.5*1.732);
      \draw[red] (-0.5,0.5*1.732)--(0.5,0.5*1.732);
      \draw (0.5,0.5*1.732)--++(0.25*1.732,-0.25);
    }{}
    \ifthenelse{\equal{#1}{lrtt}}{
      \draw (-0.05,1.732+0.5)--++(0,-0.5)--(0.55,0.55*1.732);
      \draw[red] (0.55,0.55*1.732)--(-0.55,0.55*1.732);
      \draw (-0.55,0.55*1.732)--++(-0.25*1.732,-0.25);
      \draw (0.05,1.732+0.5)--++(0,-0.5)--(-0.5,0.5*1.732);
      \draw[red] (-0.5,0.5*1.732)--(0.5,0.5*1.732);
      \draw (0.5,0.5*1.732)--++(0.25*1.732,-0.25);
    }{}
    \ifthenelse{\equal{#1}{lrtri}}{
      \draw (-0.05,1.732+0.5)--++(0,-0.5)--(-0.55,0.55*1.732)--++(-0.25*1.732,-0.25);
      \draw (0.05,1.732+0.5)--++(0,-0.5)--(0.55,0.55*1.732)--++(0.25*1.732,-0.25);
      \draw[red] (-0.45,0.5*1.732)--++(0.9,0)--++(120:0.9)--cycle;
    }{}
    \ifthenelse{\equal{#1}{lt}}{
      \draw (0,1.732+0.5)--(0,1.732)--(+0.5,0.5*1.732);
      \draw[red] (0.5, 0.5*1.732)--(-0.5,0.5*1.732);
      \draw (-0.1,1.732+0.5)--(-0.1,1.732)--(-0.55,0.55*1.732)--++(-0.25*1.732,-0.25);
      \draw (-0.5, 0.5*1.732)--++(-0.25*1.732,-0.25);
    }{}
    \ifthenelse{\equal{#1}{ltt}}{
      \draw (0,1.732+0.5)--(0,1.732)--(+0.5,0.5*1.732);
      \draw[red] (0.5, 0.5*1.732)--(-0.5,0.5*1.732);
      \draw (-0.5, 0.5*1.732)--++(-0.25*1.732,-0.25);
      \draw (-0.1,1.732+0.5)--(-0.1,1.732)--(0.35,0.55*1.732);
      \draw[red] (0.35,0.55*1.732)--(-0.55,0.55*1.732);
      \draw (-0.55,0.55*1.732)--++(-0.25*1.732,-0.25);
    }{}
    \ifthenelse{\equal{#1}{ltri}}{
      \draw (0,1.732+0.5)--(0,1.732)--(-0.5,0.5*1.732)--++(-0.25*1.732,-0.25);
      \draw[red] (-0.4,0.5*1.732)--++(0.85,0)--++(120:0.85)--cycle;
      \draw (-0.1,1.732+0.5)--(-0.1,1.732)--(-0.55,0.55*1.732)--++(-0.25*1.732,-0.25);
    }{}
    \ifthenelse{\equal{#1}{lp}}{
      \draw (0,1.732+0.5)--(0,1.732)--(-0.5,0.5*1.732)--++(-0.25*1.732,-0.25);
      \draw (-0.45,0.45*1.732)--++(60:0.9)--++(-60:0.9);
      \draw (-0.45,0.45*1.732)--++(-0.25*1.732,-0.25);
      \draw (0.45,0.45*1.732)--++(0.25*1.732,-0.25);
      \draw (-0.1,1.732+0.5)--(-0.1,1.732)--(-0.55,0.55*1.732)--++(-0.25*1.732,-0.25);
    }{}
    \ifthenelse{\equal{#1}{lpw}}{
      \draw (0,1.732+0.5)--(0,1.732)--(-0.5,0.5*1.732)--++(-0.25*1.732,-0.25);
      \draw (-0.45,0.45*1.732)--++(-0.25*1.732,-0.25);
      \draw (0.45,0.45*1.732)--++(0.25*1.732,-0.25);
      \draw[red] (-0.45,0.45*1.732)--(0.45,0.45*1.732);
      \draw (-0.1,1.732+0.5)--(-0.1,1.732)--(-0.55,0.55*1.732)--++(-0.25*1.732,-0.25);
    }{}
    \ifthenelse{\equal{#1}{lpt}}{
      \draw (0,1.732+0.5)--(0,1.732)--(+0.5,0.5*1.732);
      \draw[red] (0.5, 0.5*1.732)--(-0.5,0.5*1.732);
      \draw (-0.5, 0.5*1.732)--++(-0.25*1.732,-0.25);
      \draw (-0.45,0.45*1.732)--++(60:0.9)--++(-60:0.9);
      \draw (-0.45,0.45*1.732)--++(-0.25*1.732,-0.25);
      \draw (0.45,0.45*1.732)--++(0.25*1.732,-0.25);
      \draw (-0.1,1.732+0.5)--(-0.1,1.732)--(-0.55,0.55*1.732)--++(-0.25*1.732,-0.25);
    }{}
    \ifthenelse{\equal{#1}{lpwt}}{
      \draw (0,1.732+0.5)--(0,1.732)--(+0.5,0.5*1.732);
      \draw[red] (0.5, 0.5*1.732)--(-0.5,0.5*1.732);
      \draw (-0.5, 0.5*1.732)--++(-0.25*1.732,-0.25);
      \draw (-0.45,0.45*1.732)--++(-0.25*1.732,-0.25);
      \draw (0.45,0.45*1.732)--++(0.25*1.732,-0.25);
      \draw[red] (-0.45,0.45*1.732)--(0.45,0.45*1.732);
      \draw (-0.1,1.732+0.5)--(-0.1,1.732)--(-0.55,0.55*1.732)--++(-0.25*1.732,-0.25);
    }{}
    \node () at (-1.1,0) {};
    \node () at (1.1,0) {};
  }
}

\newcommandx{\ddwhex}[0]
{
  \draw (0.05,1.732+0.5)--++(0,-0.5-1/3*1.732)--++(1,-1/3*1.732)--++(0,-2/3*1.732)--++(-1,-1/3*1.732)--++(0,-0.5-1/3*1.732);
  \draw (-0.05,1.732+0.5)--++(0,-0.5-1/3*1.732)--++(-1,-1/3*1.732)--++(0,-2/3*1.732)--++(1,-1/3*1.732)--++(0,-0.5-1/3*1.732);
  \draw (0,0.616*1.732)--++(210:2/3*1.65)--++(0,-2/3*1.55)--(0,-0.616*1.732);
  \draw (0,0.616*1.732)--++(-30:2/3*1.65)--++(0,-2/3*1.55)--(0,-0.616*1.732);
  \fill (0,2/3*1.732-0.03) circle (0.1);
  \fill (0,-2/3*1.732+0.03) circle (0.1);
}

\newcommandx{\ddwtwostep}[1][1=]{
  \fineq[-0.4ex][0.5][1]{
    \dtptri;
    \ifthenelse{\equal{#1}{}}{
      \draw (0.05,1.732+0.5)--++(0,-0.5-1/3*1.732);
      \draw (-0.05,1.732+0.5)--++(0,-0.5-1/3*1.732);
      \draw (0.05,-1.732-0.5)--++(0,0.5+1/3*1.732);
      \draw (-0.05,-1.732-0.5)--++(0,0.5+1/3*1.732);
      \node () at (0,2/3*1.732-0.2) {$\vdots$};
      \node () at (0,-2/3*1.732+0.5) {$\vdots$};
    }{}
    \ifthenelse{\equal{#1}{ll}}{
      \draw (-0.05,1.732+0.5)--++(0,-0.5-1/3*1.732)--++(-1,-1/3*1.732)--++(0,-2/3*1.732)--++(1,-1/3*1.732)--++(0,-0.5-1/3*1.732) coordinate (A);
      \draw (0.05,1.732+0.5)--++(0,-0.5-0.05-1/3*1.732)--(0,0.616*1.732);
      \draw (0,0.616*1.732)--++(210:2/3*1.65)--++(0,-2/3*1.55)--(0,-0.616*1.732)--++(0.05,-0.05/1.732)--([shift=({0.1,0})]A);
    }{}
    \ifthenelse{\equal{#1}{lr}}{
      \draw (-0.05,1.732+0.5)--++(0,-0.5-1/3*1.732)--++(-1,-1/3*1.732)--++(0,-2/3*1.732)--++(1,-1/3*1.732)--++(0,-0.5-1/3*1.732);
      \draw (0.05,1.732+0.5)--++(0,-0.5-1/3*1.732)--++(1,-1/3*1.732)--++(0,-2/3*1.732)--++(-1,-1/3*1.732)--++(0,-0.5-1/3*1.732);
    }{}
    \ifthenelse{\equal{#1}{rl}}{
      \draw (-0.05,1.732+0.5)--++(0,-0.5-1/3*1.732)--++(1,-1/3*1.732)--++(0,-2/3*1.732)--++(-1,-1/3*1.732)--++(0,-0.5-1/3*1.732);
      \draw (0.05,1.732+0.5)--++(0,-0.5-1/3*1.732)--++(-1,-1/3*1.732)--++(0,-2/3*1.732)--++(1,-1/3*1.732)--++(0,-0.5-1/3*1.732);
    }{}
    \ifthenelse{\equal{#1}{rr}}{
      \draw (0.05,1.732+0.5)--++(0,-0.5-1/3*1.732)--++(1,-1/3*1.732)--++(0,-2/3*1.732)--++(-1,-1/3*1.732)--++(0,-0.5-1/3*1.732) coordinate (A);
      \draw (-0.05,1.732+0.5)--++(0,-0.5-0.05-1/3*1.732)--(0,0.616*1.732);
      \draw (0,0.616*1.732)--++(-30:2/3*1.65)--++(0,-2/3*1.55)--(0,-0.616*1.732)--++(-0.05,-0.05/1.732)--([shift=({-0.1,0})]A);
    }{}
    \ifthenelse{\equal{#1}{hex}}{
      \ddwhex;
    }{}
    \ifthenelse{\equal{#1}{hexa}}{
      \ddwhex;
      \node[above] () at (0,1.732+0.5) {\footnotesize $(12)(34)$};
      \node[below] () at (0,-1.732-0.5) {\footnotesize $(12)(34)$};
      \node[left] () at (-1,0) {\footnotesize $(14)(23)$};
      \node[right] () at (1,0) {\footnotesize $(24)(13)$};
    }{}
    \ifthenelse{\equal{#1}{hexb}}{
      \ddwhex;
      \node[above] () at (0,1.732+0.5) {\footnotesize $(12)(34)$};
      \node[below] () at (0,-1.732-0.5) {\footnotesize $(12)(34)$};
      \node[left] () at (-1,0) {\footnotesize $(24)(13)$};
      \node[right] () at (1,0) {\footnotesize $(14)(23)$};
    }{}
    \node () at (-1.1,0) {};
    \node () at (1.1,0) {};
  }
}

\newcommandx{\tdw}[1][1=]
{
  \fineq[-0.4ex][0.5][1]{
    \dptri;
    \ifthenelse{\equal{#1}{l}}{
      \draw (-0.1,1.732+0.5)--++(0,-0.45-1/3*1.732)--++(210:1.0);
      \draw (0,1.732+0.5)--++(0,-0.5-1/3*1.732)--++(210:1.077);
      \draw (0.1,1.732+0.5)--++(0,-0.55-1/3*1.732)--++(210:1.154);
    }{}
    \ifthenelse{\equal{#1}{r}}{
      \draw (-0.1,1.732+0.5)--++(0,-0.55-1/3*1.732)--++(-30:1.154);
      \draw (0,1.732+0.5)--++(0,-0.5-1/3*1.732)--++(-30:1.077);
      \draw (0.1,1.732+0.5)--++(0,-0.45-1/3*1.732)--++(-30:1.0);
    }{}
    \ifthenelse{\equal{#1}{llr}}{
      \draw (-0.1,1.732+0.5)--++(0,-0.45-1/3*1.732)--++(210:1.0);
      \draw (0,1.732+0.5)--++(0,-0.5-1/3*1.732)--++(210:1.077);
      \draw (0.1,1.732+0.5)--++(0,-0.45-1/3*1.732)--++(-30:1.0);
    }{}
    \ifthenelse{\equal{#1}{lrr}}{
      \draw (-0.1,1.732+0.5)--++(0,-0.45-1/3*1.732)--++(210:1.0);
      \draw (0,1.732+0.5)--++(0,-0.5-1/3*1.732)--++(-30:1.077);
      \draw (0.1,1.732+0.5)--++(0,-0.45-1/3*1.732)--++(-30:1.0);
    }{}
    \node () at (-1.1,0) {};
    \node () at (1.1,0) {};
  }
}

\newcommandx{\ugater}[4][1=0,2=0,3=,4=]{
  \begin{scope}[shift={(#1,#2)}]
      \ifthenelse{\equal{#4}{}}{
      }{
        \fill[#4] (0.75,0.25)--++(0.5,0)--++(0,1)--++(-0.5,0)--cycle;
      }
      \ifthenelse{\equal{#3}{p}}{
        \draw[line width = 2pt] (0.75,0)--(1.1,0);
        \draw (1,0)--++(0,-0.1);
      }{}
      \draw[line width = 0.5pt] (1,0)--++(0,0.25);
      \draw[line width = 0.5pt] (1,1.25)--++(0,0.25);
      \draw[line width = 0.5pt] (0.75,0.25)--++(0.5,0)--++(0,1)--++(-0.5,0);
  \end{scope}
}

\newcommandx{\ugatel}[4][1=0,2=0,3=,4=]{
  \begin{scope}[shift={(#1,#2)}]
      \ifthenelse{\equal{#4}{}}{
      }{
        \fill[#4] (0.25,0.25)--++(-0.5,0)--++(0,1)--++(0.5,0)--cycle;
      }
      \ifthenelse{\equal{#3}{p}}{
        \draw[line width = 2pt] (-0.1,0)--(0.25,0);
        \draw (0,0)--++(0,-0.1);
      }{}
      \draw[line width = 0.5pt] (0,0)--++(0,0.25);
      \draw[line width = 0.5pt] (0,1.25)--++(0,0.25);
      \draw[line width = 0.5pt] (0.25,0.25)--++(-0.5,0)--++(0,1)--++(0.5,0);
  \end{scope}
}

\newcommandx{\partitionZ}[1][1=]
{
  \foreach \x in {0,2,4}{
  	\ugate[\x][0][][][#1];
    \ugate[\x][3][][][#1];
  	\ugate[\x][6][][][#1];
  }
  \foreach \x in {1,3}{
    \ugate[\x][4.5][][][#1];
    \ugate[\x][1.5][][][#1];
  }
  \foreach \y in {1.5,4.5}{
  	\ugater[-1][\y][#1];
    \ugatel[5][\y][#1];
  }
}

\newcommandx{\ugateonlyp}[5][1=0,2=0,3=,4=,5=]{
  \begin{scope}[shift={(#1,#2)}]
      \draw[line width = 0.5pt] (0,0)--++(0,0.25);
      \draw[line width = 0.5pt] (0,1.25)--++(0,0.25);
      \draw[line width = 0.5pt] (1,0)--++(0,0.25);
      \draw[line width = 0.5pt] (1,1.25)--++(0,0.25);
      \draw[line width = 2pt] (-0.1,0)--(1.1,0);
      \draw (0,0)--++(0,-0.1);
      \draw (1,0)--++(0,-0.1);
  \end{scope}
}

\newcommandx{\ugateronlyp}[3][1=0,2=0,3=]{
  \begin{scope}[shift={(#1,#2)}]
    \draw[line width = 2pt] (0.75,0)--(1.1,0);
    \draw (1,0)--++(0,-0.1);
  \end{scope}
}

\newcommandx{\ugatelonlyp}[3][1=0,2=0,3=]{
  \begin{scope}[shift={(#1,#2)}]
    \draw[line width = 2pt] (-0.1,0)--(0.25,0);
    \draw (0,0)--++(0,-0.1);
  \end{scope}
}

\newcommandx{\partitionZc}[1][1=]
{
  \foreach \x in {0,...,5}{
  	\draw (\x, 0)--++(0,7.5);
  }
  \foreach \y in {0,...,4}{
  	\fill[black!40] (-0.25,1.25+\y*1.5)--++(5.5,0)--++(0,-1)--++(-5.5,0)--cycle;
    \draw (-0.25,1.25+\y*1.5)--++(5.5,0);
    \draw (-0.25,0.25+\y*1.5)--++(5.5,0);
  }
}

\newcommandx{\partitionZtwosite}[1][1=]
{
  \partitionZc
  \foreach \x in {0,2,4}{
  	\ugateonlyp[\x][0][][];
    \ugateonlyp[\x][3][][];
  	\ugateonlyp[\x][6][][];
  }
  \foreach \x in {1,3}{
    \ugateonlyp[\x][4.5][][];
    \ugateonlyp[\x][1.5][][];
  }
  \foreach \y in {1.5,4.5}{
  	\ugateronlyp[-1][\y];
    \ugatelonlyp[5][\y];
  }
}

\newcommandx{\partitionZonesite}[1][1=]
{
  \partitionZc
  \foreach \x in {0,...,5}{
   	\foreach \y in {0,1.5,3,4.5,6}{
      \begin{scope}[shift={(\x,\y)}]
        \draw[line width = 2pt] (-0.25,0)--(0.25,0);
      \end{scope}
    }
  }
}

\newcommandx{\threegate}[1][1=]
{\fineq[-0.8ex][0.4][0.8]{
  \ugate[-1][1.5][][$\sigma_b$];
  \ugate[1][1.5][][$\sigma_c$];
  \ugate[0][0][][$\sigma_a$];
}
}

\newcommandx{\ptensor}[2][1=,2=]
{
  \fineq[-0.8ex][0.35][0.8]{
    \sqzbox[0][0][#2][$#1$];
    \draw (0.5,0)--++(0,-0.25);
    \draw (1.75,0)--++(0,-0.25);
    \draw (0.5,1)--++(0,0.25);
    \draw (1.75,1)--++(0,0.25);
  }
}

\newcommandx{\pdecomp}[8][1=0,2=0,3=,4=,5=,6=,7=,8=]
{
  \begin{scope}[shift={(#1,#2)}]
    \drawbox[0][0][1][0.4][#6][$#5$];
    \drawbox[1.25][0][2.25][0.4][#8][$#7$];
    \drawbox[0][0.6][2.25][1][#4][$#3$];
    \draw (0.5,0)--++(0,-0.25);
    \draw (1.75,0)--++(0,-0.25);
    \draw (0.5,1)--++(0,0.25);
    \draw (1.75,1)--++(0,0.25);
  \end{scope}
}

\newcommandx{\pdecompeq}[8][1=0,2=0,3=,4=,5=,6=,7=,8=]
{
   \fineq[-0.8ex][0.5][0.55]{
     \pdecomp[#1][#2][#3][#4][#5][#6][#7][#8]
  }
}

\newcommandx{\hfpdecomp}[5][1=0,2=0,3=,4=,5=]
{
  \begin{scope}[shift={(#1,#2)}]
    \drawbox[0][0][1][0.4][#5][$#4$];
    \draw (0.5,0)--++(0,-0.25);
    \draw (0.5,1)--++(0,0.25);
    \ifthenelse{\equal{#3}{l}}{
      \draw (0,0.6)--++(1,0)--++(0,0.4)--++(-1,0);
    }{}
    \ifthenelse{\equal{#3}{r}}{
      \draw (1,0.6)--++(-1,0)--++(0,0.4)--++(1,0);
    }{}
  \end{scope}
}

\newcommandx{\lshapetensor}[4][1=,2=,3=l,4=]{
  \fineq[-0.8ex][0.5][0.55]{
    \drawbox[0][-0.65][2.25][-0.25][#4][$#2$];
    \draw (0.5,0)--++(0,-0.25);
    \draw (2.25-0.5,0)--++(0,-0.25);
    \ifthenelse{\equal{#3}{l}}{
      \drawbox[0][0][1][0.4][#4][$#1$];
    }{
      \drawbox[1.25][0][2.25][0.4][#4][$#1$];
    }
  }
}

\newcommandx{\hfpdecompeq}[6][1=0,2=0,3=,4=,5=,6=]
{
   \fineq[-0.8ex][0.35][0.8]{
     \hfpdecomp[#1][#2][#3][#4][#5][#6];
  }
}

\newcommandx{\hexboxdecomp}[8][1=,2=,3=,4=,5=,6=,7=,8=]
{
  \fineq[-0.8ex][0.5][0.55]{
    \pdecomp[0][2.5][][][][][+][blue!50];
    \pdecomp[2.5][2.5][][][-][blue!50];
    \hfpdecomp[0][1.25][l][+][green!50];
    \sqzbox[1.25][1.25][red!50][\Huge $\perp$][#8];
    \hfpdecomp[1.5+2.25][1.25][r][-][green!50];
    \pdecomp[0][0][\sigma_1][green!50];
    \pdecomp[2.5][0][\sigma_2][green!50];
  }
}

\newcommandx{\triboxdecomp}[8][1=,2=,3=,4=,5=,6=,7=,8=]
{
  \fineq[-0.8ex][0.5][0.55]{
    \hfpdecomp[0][1.25][l][#1][red!50];
    \hfpdecomp[1.25][1.25][r][#2][red!50];
    \pdecomp[0][0][#3][red!50];
  }
}

\newcommandx{\partitionZs}[1][1=]
{
  \ugate[0][0][][$s_{ x-2,t-2}$];
  \ugate[1][1.5][][$s_{ x-1,t-1}$];
  \ugate[0][3][][$s_{ x-2,t}$];
  \ugate[1][4.5][][$s_{ x-1,t+1}$];
  \ugate[0][6][][$s_{ x-2,t+2}$];

  \ugate[2][0][][$s_{ x,t-2}$];
  \ugate[4][0][][$s_{ x+2,t-2}$];
  \ugate[3][1.5][][$s_{ x+1,t-1}$];
  \ugate[2][3][][$s_{ x,t}$];
  \ugate[4][3][][$s_{ x+2,t}$];
  \ugate[3][4.5][][$s_{ x+1,t+1}$];
  \ugate[2][6][][$s_{ x,t+2}$];
  \ugate[4][6][][$s_{ x+2,t+2}$];
  \ugater[-1][1.5][];
  \ugater[-1][4.5][];
  \ugatel[5][1.5][];
  \ugatel[5][4.5][];
}

\newcommandx{\filled}[1][1=0]
{
  \fineq[-0.5ex]{\fill (0,0) circle (0.08);}
}

\newcommandx{\hket}[2][1=01,2=]
{
  \ifthenelse{\equal{#2}{}}{
    |
  }{}
  \fineq[-0.7ex]{
    \ifthenelse{\equal{#1}{0}}{
      \draw (0,0) circle (0.08);
    }{}
    \ifthenelse{\equal{#1}{1}}{
      \fill (0,0) circle (0.08);
    }{}
    \ifthenelse{\equal{#1}{00}}{
      \draw (0,0) circle (0.08);
      \draw (0.3,0) circle (0.08);
    }{}
    \ifthenelse{\equal{#1}{01}}{
      \draw (0,0) circle (0.08);
      \fill (0.3,0) circle (0.08);
    }{}
    \ifthenelse{\equal{#1}{10}}{
      \fill (0,0) circle (0.08);
      \draw (0.3,0) circle (0.08);
    }{}
    \ifthenelse{\equal{#1}{11}}{
      \fill (0,0) circle (0.08);
      \fill (0.3,0) circle (0.08);
    }{}
  }
  \rangle    
}